\documentclass{article}
\usepackage[utf8]{inputenc}
\usepackage{authblk}
\usepackage{setspace}
\usepackage[margin=1.25in]{geometry}
\usepackage{graphicx}
\graphicspath{ {./} }
\usepackage{subcaption}
\usepackage{amsmath}

\usepackage{algorithm}
\usepackage{algorithmicx}
\usepackage{algpseudocode}
\usepackage{amssymb}
\usepackage{cases}
\usepackage{float}
\usepackage{makecell}
\usepackage{epstopdf}
\usepackage{multirow}
\usepackage{pifont}

\newtheorem{definition}{Definition}[section]

\floatname{algorithm}{Algorithm}

\usepackage[style=nejm, citestyle=numeric-comp, sorting=none]{biblatex}
\title{Differentiated Service Entanglement Routing for Quantum Networks}

\author[1,2$\dag$]{Hui Han}
\author[2*$\dag$]{Bo Liu}
\author[1,3]{Bang-Ying Tang}
\author[2,4]{Si-Yu Xiong}
\author[2,5]{Jin-Quan Huang}
\author[1*]{Wan-Rong Yu}
\author[1]{Shu-Hui Chen}

\affil[1]{College of Computer, National University of Defense Technology, Changsha 410073, China.}
\affil[2]{College of Advanced Interdisciplinary Studies, National University of Defense Technology,
Changsha 410073, China.}
\affil[3]{Strategic Assessments and Consultation Institute, Academy of Military Sciences, Beijing 100091, China.}
\affil[4]{School of Mathematical Sciences, Sichuan Normal University, Chengdu 610068, China.}
\affil[5]{School of Electronics and Communication Engineering, Shenzhen Campus of Sun Yat-sen University, Shenzhen, 518107, China.}
\affil[*]{Address correspondence to: wlyu@nudt.edu.cn (W.-R.Y.); liubo08@nudt.edu.cn (B.L.)}
\affil[$\dag$]{These authors contributed equally to this work.}

\date{}

\begin{document}

\maketitle

\captionsetup[figure]{labelfont={bf},name={Fig.},labelsep=period}

\begin{abstract}
    The entanglement distribution networks with various topologies are mainly implemented by active wavelength multiplexing routing strategies. However, designing an entanglement routing scheme, which achieves the maximized network connections and the optimal overall network efficiency simultaneously, remains a huge challenge for quantum networks. In this article, we propose a differentiated service entanglement routing (DSER) scheme, which firstly finds out the lowest loss paths and supported wavelength channels with the tensor-based path searching algorithm, and then allocates the paired channels with the differentiated routing strategies. The evaluation results show that the proposed DSER scheme can be performed for constructing various large scale quantum networks.
\end{abstract}


\section{Introduction}
\label{chap:Introduction}
The quantum network extends the capabilities of classical networks, enabling a range of applications impossible for classical networks, such as quantum key distribution~\cite{weiHighSpeedMeasurementDeviceIndependentQuantum2020,fan-yuanRobustAdaptableQuantum2022}, distributed quantum computing~\cite{yinQuantumTeleportationEntanglement2012,krauterDeterministicQuantumTeleportation2013}, quantum clock synchronization~\cite{quanImplementationFieldTwoway2022,hongDemonstration50Km2022} and distributed quantum sensing~\cite{guoDistributedQuantumSensing2020,jiangQuantumSensingRadiofrequency2023}. The research on quantum networks has been stepped from the trusted relayed network stage into the quantum entanglement distribution network stage~\cite{elliottCurrentStatusDARPA2005,dianatiArchitectureSecoqcQuantum2007,sasakiFieldTestQuantum2011,chenIntegratedSpacetogroundQuantum2021,liuPhotonicreconfigurableEntanglementDistribution2023}.

In 2018, the first fully connected entanglement network was realized by wavelength division multiplexing (WDM) strategy, however thiskind network scale would be limited due to $O(n^2)$ wavelength channels are required for $n$ distant users~\cite{wengerowskyEntanglementbasedWavelengthmultiplexedQuantum2018}. Afterwards, a linear complexity entanglement network was implemented in 2020, which combined the WDM and beam splitting strategies~\cite{joshiTrustedNodeFree2020}. Recently, various configurable entanglement networks were demonstrated with wavelength selective switches, which can satisfy dynamic requests of users with adaptive bandwidth management schemes~\cite{alshowkanReconfigurableQuantumLocal2021,appasFlexibleEntanglementdistributionNetwork2021}.

Due to the limited entanglement bandwidth and connection paths in the quantum network, an efficient entanglement routing scheme is always required when considering the maximized network connections and the optimal overall network
efficiency simultaneously~\cite{banerjeePracticalApproachRouting1996,ozdaglarRoutingWavelengthAssignment2003,leDQRADeepQuantum2022,husseinReviewVariousQuantum2022,santosShortestPathFinding2023}. In 2021, Lingaraju \textit{et al.} show the adaptively partitioning bandwidth with a single wavelength-selective switch, but without the design of dynamic routing algorithms~\cite{lingarajuAdaptiveBandwidthManagement2021}. In 2022, Li \textit{et al.} proposed a first request first service routing scheme aimed at providing users the lowest loss path when presetting a single channel~\cite{liFirstRequestFirst2022}. However, this scheme only achieves the maximum connections without considering the optimal application performance of users.
 
The majority proposed quality of service (QoS) routing schemes for quantum networks were mainly implemented with trusted nodes~\cite{mehicNovelApproachQualityofService2020,parkQualityServiceEvaluation2022}. Currently, QoS routing has also been started to be considered in the entanglement distribution network stage with strategies such as genetic algorithm and application priority ranking~\cite{alnasOptimalResourceAllocation2022,chenAPRQKDNQuantumKey2022}. However, almost all QoS routing schemes are driven on the performance of the certain single-user request, which are not suitable for the quantum network providing QoS routing service while keeping the largest active entanglement connections~\cite{saxenaDifferentiatedServicesIts2021,cicconettiQualityServiceQuantum2022a,mehicQualityServiceArchitectures2022}. 

In this article, we propose a differentiated service entanglement routing scheme (DSER) for quantum entanglement distribution networks, which allocates the optimal wavelength channels for multi-users with different QoS strategies. Firstly, the DSER scheme can find the lowest-loss path for users with all potential supported channels by performing the modified tensor-based Dijkstra algorithm. Then, the first request first routing strategy is implemented in the DSER scheme to allocate the optimal channels for distant users with the differentiated QoS cases. Finally, the evaluation results show the robustness and efficiency of the proposed DSER scheme for constructing quantum entanglement networks with various definable topologies.

\section{Modeling of the quantum entanglement distribution network}
\label{chap:Model}

\begin{figure}[ht]
 \centering
 \includegraphics[width=1\linewidth]{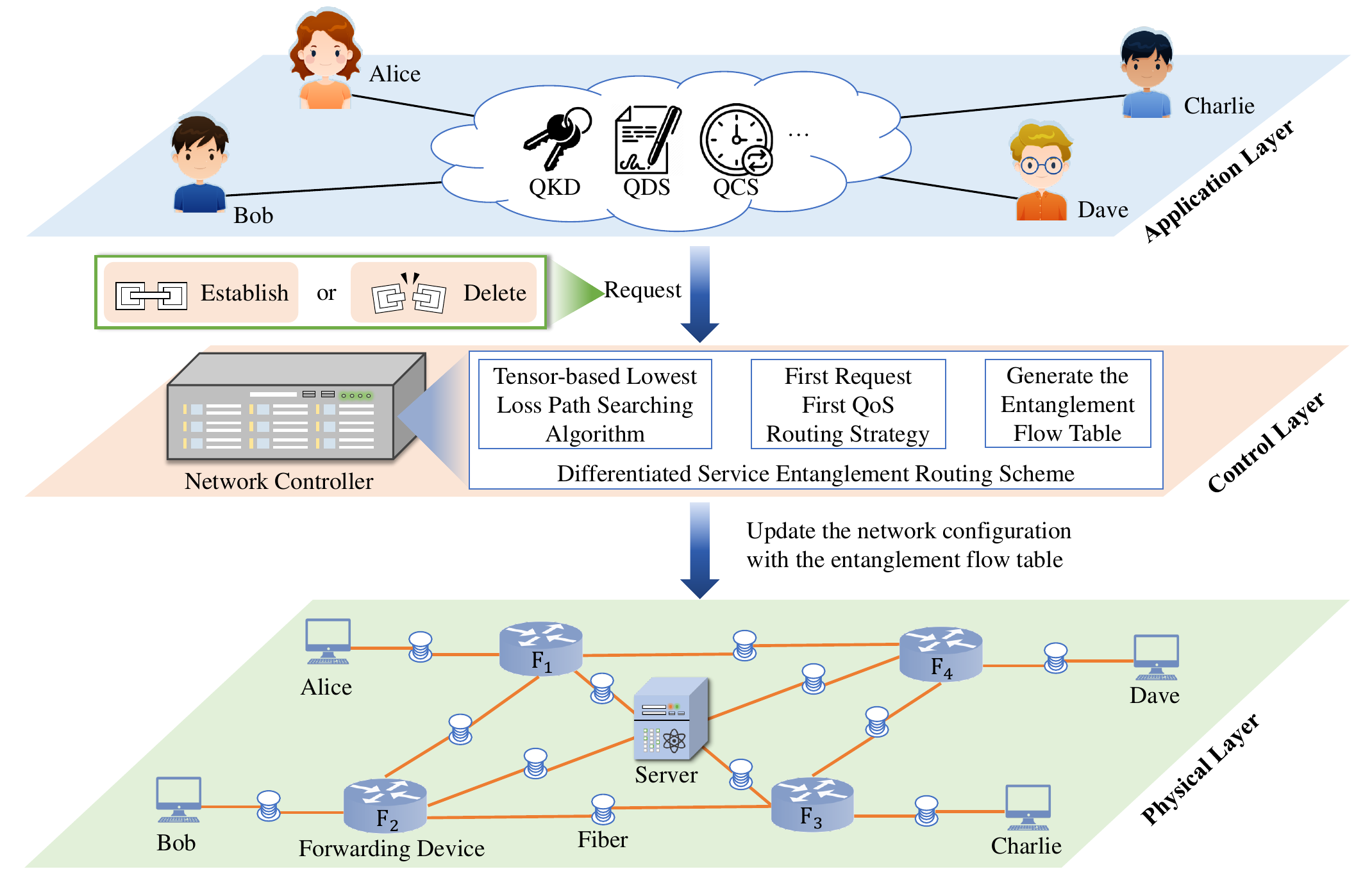}
 \caption{The architecture of the configurable quantum entanglement distribution network. The network can be divided into the physical layer, the control layer and the application layer. The user's request for establishing or deleting entanglement connections will be transmitted to the control layer. The controller performs the routing strategies according to different user requirements and generates the entanglement flow tables to the physical layer. Afterwards, the physical layer configures the forwarding devices according to the entanglement flow tables. QKD is short for quantum key distribution, QDS is short for quantum digital signature and QCS is short for quantum clock synchronization.}
 \label{pic: model}
\end{figure}

Due to the limited entanglement resources, it is difficult for static networks to support constructing large-scale entanglement distribution networks. Meanwhile, the current entanglement network cannot automatically address, store, and forward quantum signals. Therefore, the architecture of the configurable entanglement distribution network is focused on in this article, as shown in Figure~\ref{pic: model}.

\begin{enumerate}
	\item \textbf{Physical Layer}, 
    which generates, transmits, and detects the quantum entangled photons, mainly involving a quantum server composed of an entangled photon source, forwarding devices, and user terminals, where forwarding devices include wavelength selective switches, dense wavelength division multiplexers, and optical switches. The user terminals are usually implemented to decode, detect, and process entangled photons according to different protocol requirements. All optical components are connected via optical fiber or free-space quantum links.

	\item \textbf{Control Layer},
    which calculates and generates the entanglement flow tables, mainly involving a quantum network controller. The controller develops routing strategies according to different user requirements and generates the entanglement flow tables to configure forwarding devices. This article especially focuses on proposing a differentiated service entanglement routing scheme, including a tensor-based lowest loss path searching algorithm, first request first quality of service strategy, and generation of the entanglement flow tables, which is described in detail in the Section.~\ref{chap: Methods}.

    \item \textbf{Application Layer},
    which implements the quantum key distribution, quantum digital signature, quantum clock synchronization, and other protocols according to the user's application requirements. 
    The user's request for establishing or deleting entanglement connections will be transmitted to the control layer. These protocols have different requirements on network optimal entanglement resources, and quantum network requires differentiated service to meet the user's request in different cases.

\end{enumerate}

Most quantum network structures are modeled as a graph $G=\langle V, E\rangle $~\cite{leiferQuantumGraphicalModels2008,kobayashiConstructingQuantumNetwork2011,eppingRobustEntanglementDistribution2016,meignantDistributingGraphStates2019,azumaToolsQuantumNetwork2021,liFirstRequestFirst2022}.
The quantum information processing nodes in the quantum network are considered as the nodes $V$, and the quantum channels (fiber or free-space link) that send quantum information are considered the edge $E$. The previous work has defined the network in our group. 

First, $V$ is the set of $k$ nodes. Each node $\nu_i$ contains a pass-through loss tensor and a support channel vector, defined as a bi-variate vector $\nu_i\triangleq(\mathcal{W}_i,T_i), 1\leq i\leq k$. $\mathcal{W}_i\in \mathbb{R}^{m_i\times m_i\times |T_i|}$, $m_i$ is the total counts of ports for node $\nu_i$ and $T_i$ is the supported wavelength channel vector, consisting of the channel indexes defined by the International Telecommunication Union (ITU) Grid C-Band (100 GHz Spacing) standard. The size of $T_i$ shows the divided number of supported wavelength channels of $\nu_i$.
Second, an element $e^{ij}_{\alpha \beta}$ in set $E$ is an edge from the $\alpha$-th port of node $\nu_i$ to another port $\beta$-th of node $\nu_j$, defined as a five-variable vector $e^{ij}_{\alpha \beta}=(i,\alpha,j,\beta,c^{ij}_{\alpha \beta})$, where $1\leq i,j\leq k, 1\leq \alpha\leq m_i, 1\leq \beta\leq m_j$. 
The $e^{ij}_{\alpha \beta}[\gamma]$ represents the $\gamma$-th element in $e^{ij}_{\alpha \beta}$, $\gamma\in\{1,2,3,4,5\}$ and $e^{ij}_{\alpha \beta}[5]$ is the path loss (see the Appendix~\ref{chap: Modeling of the network}).

\section{Methods} 
\label{chap: Methods}

Designing an entanglement routing and wavelength assignment scheme for quantum networks remains a huge challenge when attempts to achieve the maximized network connections and the optimal overall network efficiency simultaneously. To address this issue, we propose a novel differentiated service entanglement routing (DSER) scheme, which firstly finds out the lowest-loss paths and supported wavelength channels with the tensor-based path searching algorithm, and then allocates the optimal entanglement resources for multi-users with the differentiated routing strategies. Meanwhile, in order to analysis the optimal performance of the proposed DSER scheme, we suppose the BBM92 quantum key distribution protocol is performed for the quantum networks. 
 
\subsection{the Proposed DSER Scheme}

\begin{figure}
    \centering
    \includegraphics[width=1\linewidth]{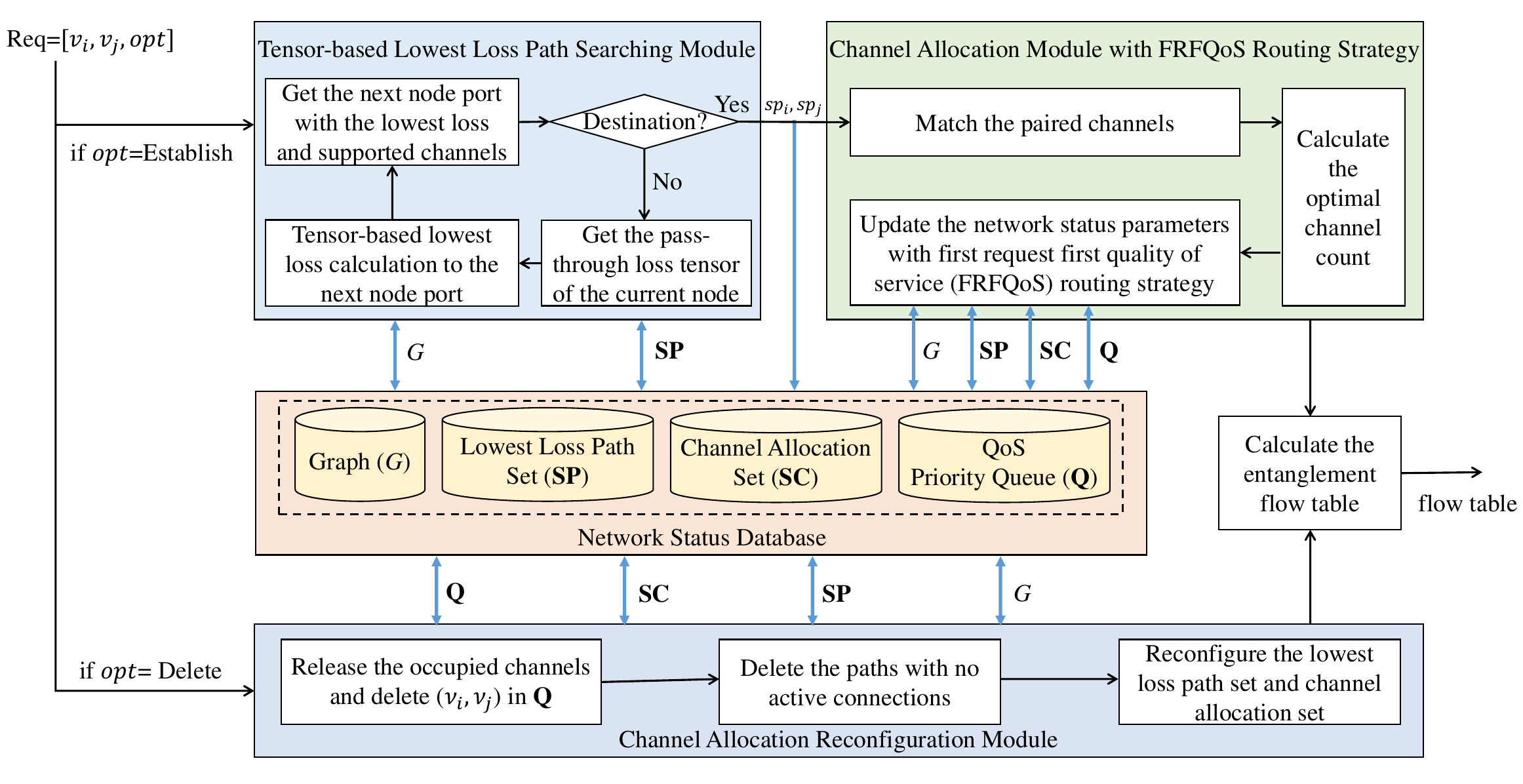}
    \caption{Schematic diagram of the DSER scheme. $\mathrm{Req}=[\nu_i,\nu_j,opt]$ means the routing request between user $\nu_i$ and $\nu_j$ and $opt\in [\mathrm{Establish}, \mathrm{Delete}]$. The network graph $G$, the lowest loss path set \textbf{SP}, the channel allocation set \textbf{SC} and the QoS priority queue \textbf{Q} are stored in the network status database. The blue arrow indicates the data transmission lines and the black arrow indicates the flow control lines.}
    \label{pic:routing}
\end{figure}

The schematic diagram of the proposed DSER scheme is shown in Fig.~\ref{pic:routing}, which mainly including the tensor-based lowest loss path searching module, the channel allocation module with first request first quality of service (FRFQoS) routing strategy, and the channel allocation reconfiguration module. The network graph $G$, the lowest loss path set \textbf{SP}, the channel allocation set \textbf{SC} and the QoS priority queue \textbf{Q} are stored in the network status database, which is updated after each module. Afterwards, the entanglement flow table can be calculated and distributed to the network forwarding devices.

First, in order to describe the DSER scheme clearly, the definitions of $\textbf{SP}$, $\textbf{SC}$ and $\textbf{Q}$ are given as follows:
\begin{definition}
    The lowest loss path set \rm{\textbf{SP}}$=\{sp_1,sp_2,\cdots,sp_{n_1}\}$, where $n_1$ is the size of $\rm{\textbf{SP}}$ and $sp_i=(p_i,l_i,c^a_i,c^o_i)$, $1\leq i \leq n_1$, $sp_i[\gamma_1]$ represents the $\gamma_1$-th element in $sp_i$, $\gamma_1\in\{1,2,3,4\}$. Here, $p_i$ is the found-out lowest-loss path from the entangled photon source to the network node $\nu_i$, $l_i$ is overall loss value of the path $p_i$, $c^\mathrm{a}_i$ ($c^\mathrm{o}_i$) is the available (occupied) channel vector of the path $p_i$.
    \label{The lowest loss path set}
\end{definition}

\begin{definition}
    The channel allocation set \rm{\textbf{SC}}$=\{sc_1,sc_2,\cdots,sc_{n_2}\}$, where $n_2$ is the size of \rm{\textbf{SC}} and $sc_i=(ind,\nu_q,\nu_r,tag_{qr})$, $1\leq i \leq n_2$, $sc_i[\gamma_2]$ represents the $\gamma_2$-th element in $sc_i$, $\gamma_2\in\{1,2,3,4\}$. Here, $ind$ is the wavelength channel index supported in the whole quantum network, which is occupied by the user $\nu_q$ and the entangled paired user $\nu_r$, with the requested timetag value of $tag_{qr}$.
    \label{The channel allocation set}
\end{definition}

\begin{definition}
    The priority queue \rm{\textbf{Q}}$=\{q_1,q_2,\cdots,q_{n_3}\}$, where $n_3$ is the size of \rm{\textbf{Q}} and $q_i=(\nu_q,\nu_r,$ $lk_{qr},tag_{qr})$, $1\leq i \leq n_3$, $q_i[\gamma_3]$ represents the $\gamma_3$-th element in $q_i$, $\gamma_3\in\{1,2,3,4\}$. $lk_{qr}=1$ indicates the entanglement connection has been established between the paired node $\nu_q$ and $\nu_r$, else $lk_{qr}=0$. $q_i$ has the highest priority if $tag_{qr}$ is the minimum value under the condition of $lk_{qr}=0$. Meanwhile, $q_i$ has the lowest priority if $tag_{qr}$ is the maximum value under the condition of $lk_{qr}=1$.
    \label{The priority queue}
\end{definition}

Then, we will describe the DSER scheme detailly under two different scenarios defined by the routing request $\mathrm{Req}=[\nu_i,\nu_j,opt]$.

\textbf{Case 1:} $opt=\mathrm{Establish}$, which means establish an entanglement connection between the distant user $\nu_i$ and $\nu_j$.

Firstly, for node $\nu_i$ ($\nu_j$), calculate $sp_i$ ($sp_j$) with the tensor-based lowest loss path searching module. If $sp_i \in \textbf{SP}$ ($sp_j \in \textbf{SP}$), the $sp_i$ ($sp_j$) can be directly readout from the set $\textbf{SP}$. Else, obtain the next node port with the lowest loss and supported channels from the entangled photon source, if the target node $\nu_i$ ($\nu_j$) is not found, get the pass-through loss tensor of the current node. Afterwards, perform the tensor-based lowest loss calculation procedure to the next node port. Then, repeats this process until the next node is the destination $\nu_i$ ($\nu_j$). Finally, lock the internal ports of the time division multiplexing (TDM) forwarding devices on the path $p_i$ by the operation $V=Lock_{port}(G,p_i)$, which is given in Definition~\ref{def:lockport} (Appendix~\ref{chap:lock and unlock}).

Secondly, calculate set $\textbf{SC}$ with the channel allocation module by performing the first request first QoS (FRFQoS) routing strategy. Find out all matched paired channels with $sp_i$ and $sp_j$. Afterwards, calculate the optimal channel count and update the network status parameters with the FRFQoS routing strategy, which is detailed described in Section~\ref{chap: first request first QoS (FRFQoS) routing strategy}. Meanwhile, the matched channel vector $ch_i$ and $ch_j$ of the node $\nu_i$ and $\nu_j$ are calculated from the set $\textbf{SC}$. Finally, lock the paired channel of the internal ports for the WDM forwarding devices on the path $p_i$ and $p_j$ by the operation $V=Lock(G,p_x,ch_x)$, where $x =i,j$ (see the Appendix~\ref{chap:lock and unlock} Definition~\ref{def:lock}).

Thirdly, calculate and distribute the entanglement flow table to the forwarding devices (see the Appendix~\ref{chap: Entanglement flow table}).

\textbf{Case 2:} $opt=\mathrm{Delete}$, which means delete an entanglement connection between the distant users.

Firstly, release the occupied paired channel vector $ch_i$ and $ch_j$ for $(\nu_i,\nu_j)$ in set $\textbf{SC}$ and delete the user pair $(\nu_i,\nu_j)$ if they are in queue $\textbf{Q}$. 

Secondly, delete the paths with no active connections in the set $\textbf{SP}$, with the unlock operation $V=Unlock_{port}(G,p_x)$ and $V=Unlock(G,p_x,ch_x)$, where $x =i,j$ (see the Appendix~\ref{chap:lock and unlock} Definition~\ref{def:unlockport} and~\ref{def:unlock}).

Thirdly, Reconfigure the lowest loss path set $\textbf{SP}$ and the channel allocation set $\textbf{SC}$ according to the queue $\textbf{Q}$. If there are user pairs with $lk=0$, search the path information by the tensor-based lowest loss path searching algorithm and allocate the channels with the differentiated QoS cases. Else, allocate the matched channels to the user pairs with $lk=1$. Meanwhile, the user pair who satisfy the QoS requirements will leave the queue $\textbf{Q}$. 

Fourthly, calculate and distribute the entanglement flow table to the forwarding devices.

\subsection{Tensor-based lowest loss path searching algorithm}
\label{chap: Tensor-based lowest loss path searching algorithm}

The performance criteria of entanglement distribution applications are usually mainly affected by the link loss $l_i$ ($l_j$) from the entangled photon source to node $\nu_i$ (node $\nu_j$). Due to limited wavelength and time multiplexing resources, finding out the optimal paired paths efficiently from the entangled photon source to multiple distant users is still a huge challenge. To solve this problem, based on our previous work~\cite{liFirstRequestFirst2022}, we propose a novel \textbf{Te}nsor-based \textbf{l}owest \textbf{l}oss \textbf{p}ath \textbf{s}earching (shorted as Tell-PS) algorithm, which can find out the optimal path and supported wavelength channels between the requested node and the entangled photon source. As shown in Fig.~\ref{pic: path}, the proposed Tell-PS algorithm mainly consists of eight steps, which can be described as followings:
\begin{figure}
    \centering
    \includegraphics[width=1\linewidth]{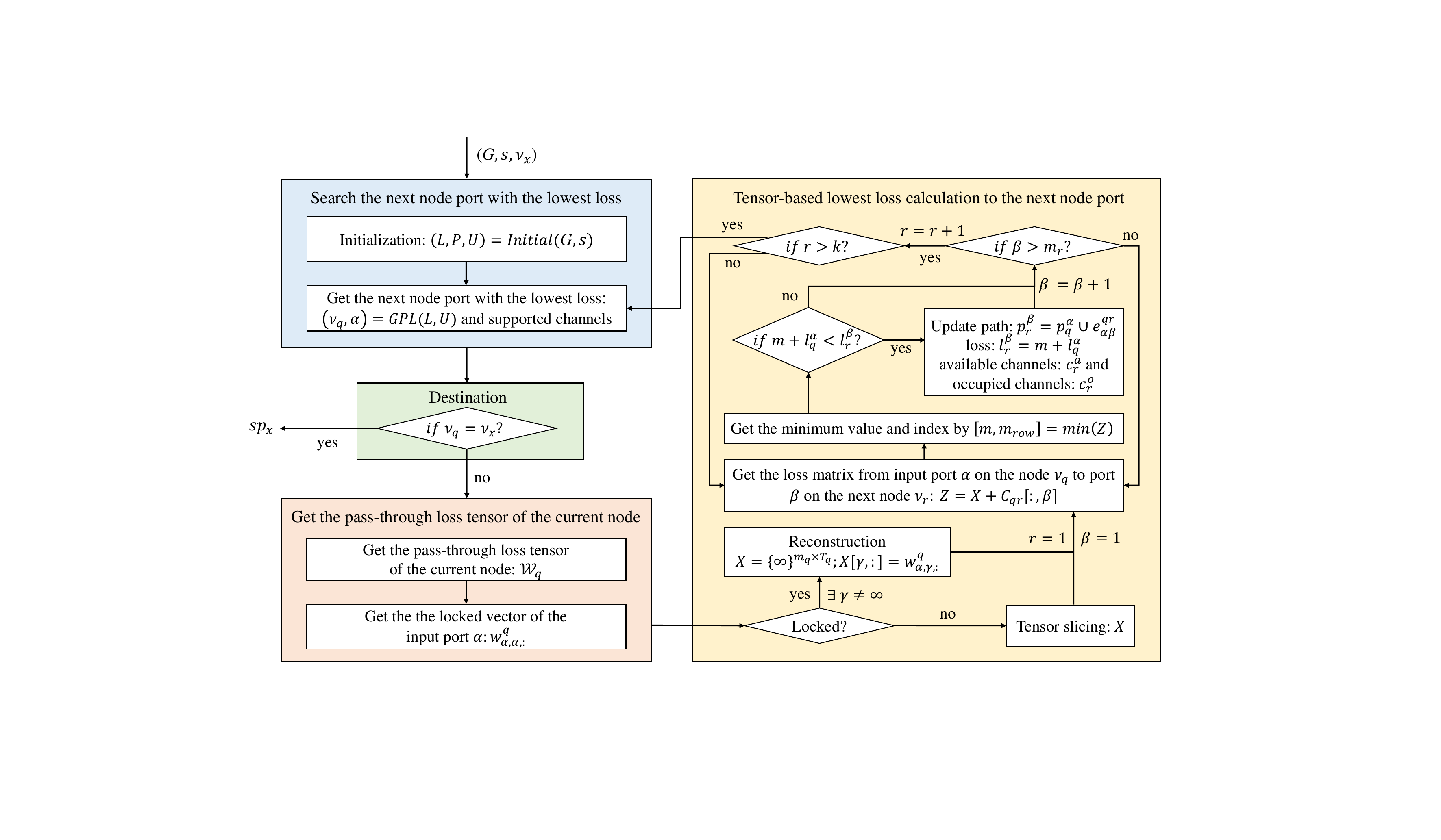}
    \caption{Schematic diagram of the proposed Tell-PS algorithm. $L=\{l_r^\beta\}, P=\{p_r^\beta\}, U=\{u_r^\beta\}, r=1,2,\cdots,k$ and $\beta=1,2,\cdots,m_r$, $k$ is the total count of nodes, $m_r$ is the total supported port count of the node $\nu_r$.  $l_r^\beta$ is the link loss between $\nu_s$ and the $\beta$-th port of $\nu_r$, $p_r^\beta$ is a vector composed of the edge from $\nu_s$ to the $\beta$-th port of $\nu_r$. If the $\beta$-th port of $\nu_r$ is not directly connected to the node $\nu_s$, $p_r^\beta=\emptyset$ and $l_r^\beta=\infty$. $u_r^\beta$ represents the visit status of the $\beta$-th port of $\nu_r$.
    $(\nu_q,\alpha)=GPL(L,U)$ returns the $\alpha$-th port of next node $\nu_q$ with the lowest loss. $w_{\alpha,\alpha,:}^q$ represents the $\alpha$-th port input and $\alpha$-th port output loss vector of pass-through loss tensor $\mathcal{W}_q$ for node $\nu_q$. $C_{qr}[:, \beta]$ represents the $\beta$-th column of $C_{qr}$. $[m,m_{row}]=min(Z)$ represents getting the minimum value $m$ and index number $m_{row}$ in $Z$.}
    \label{pic: path}
\end{figure}

\textbf{Step 1.} Initialize the network with $(L,P,U)=Initial(G,s)$, where $L=\{l_r^\beta\}, P=\{p_r^\beta\}, U=\{u_r^\beta\}, r=1,2,\cdots,k$ and $\beta=1,2,\cdots,m_r$. $k$ is the total count of nodes, $m_r$ is the total supported port count of the node $\nu_r$. $l_r^\beta$ is the link loss between $\nu_s$ and the $\beta$-th port of $\nu_r$. $p_r^\beta$ is a vector composed of the edge from $\nu_s$ to the $\beta$-th port of $\nu_r$. If the $\beta$-th port of $\nu_r$ is not directly connected to the node $\nu_s$, $p_r^\beta=\emptyset$ and $l_r^\beta=\infty$. $u_r^\beta$ represents the visit status of the $\beta$-th port of $\nu_r$, where ``0'' means ``unvisited'' and ``1'' means ``visited''. $u_s^\beta$ is initialized to ``1'' and other element of $U$ are set as ``0'', $\beta=1,2,\cdots,m_s$. Meanwhile, initialize the lowest loss path set $\textbf{SP} = \emptyset$.

\textbf{Step 2.} Find out the next unvisited node $\nu_q$ port $\alpha$ with the lowest loss with the operation $(\nu_q,\alpha)=GPL(L,U)$. Meanwhile, update $u_{q}^{\alpha}=1$.

\begin{definition}
	$(\nu_q,\alpha)=GPL(L,U)$. For $\forall$ $\delta \in \left\{1,2,\cdots,k\right\} $ and $\varphi \in \left\{1,2,\cdots,m_\delta\right\} $, find the next unvisited node $\nu_q$ and its port index $\alpha$, which holds $l^\alpha_q$ = $min\left\{l^\varphi _\delta +u^\varphi _\delta \times \infty\right\}$, where $L=\{l_\delta^\varphi\}$ and $U=\{u_\delta^\varphi\}$.
\end{definition}

\textbf{Step 3.} If $\nu_q = \nu_x$, output the element $sp_x$ and stop. Else, get the pass-through loss tensor $\mathcal{W}_q$ of the current node $\nu_q$ and get the locked vector $w_{\alpha,\alpha,:}^q$ of the input port $\alpha$ in $\mathcal{W}_q$.

\textbf{Step 4.} If $w_{\alpha,\alpha,:}^q=\infty$, show that the $\alpha$-th port of node $\nu_q$ is not locked, then perform the tensor slicing operation to extract the loss matrix $X=W_{\alpha,:,:}^q$ of the input port $\alpha$. Otherwise, $\exists \gamma \neq \infty, \gamma \in w_{\alpha,\alpha,:}^q$, construct $X=\{\infty\}^{m_q\times T_q}$ and then update $X[\gamma,:]=\mathrm{w}_{\alpha,\gamma,:}^q$. Afterwards, set $r=1$ and $\beta=1$.

\textbf{Step 5.}  Calculate the total loss from the input port $\alpha$ of $\nu_q$ to port $\beta$ of node $\nu_r$ by $Z=X+C_{qr}[:,\beta]$ with size of $m_q \times T_q$, where $C_{qr}[:,\beta]$ is the edge loss, represented by the $\beta$-th column of $C_{qr}$. 

\textbf{Step 6.} Get the minimum value and the row index by $[m,m_{row}]=min(Z)$. If $m+l_q^\alpha<l_r^\beta$, update the path vector by $p_r^\beta=p_q^\alpha \cup e_{\alpha \beta}^{qr}$, update the loss value by $l_r^\beta=m+l_q^\alpha$, update the available channels by $c_r^a = \{Z(m_{row},:)\neq \infty\}$ and update the occupied channels by $c_r^o = \{Z(m_{row},:)= \infty\}$. Else, update the port index by $\beta=\beta+1$.

\textbf{Step 7.} If $\beta>m_r$, update the node index $r=r+1$. Else, back to \textbf{Step 5}.

\textbf{Step 8.} If $r>k$, back to \textbf{Step 2}. Otherwise, set $\beta=1$ and back to \textbf{Step 5}.

\subsection{First request first QoS (FRFQoS) entanglement routing strategy} 
\label{chap: first request first QoS (FRFQoS) routing strategy}

Here, we propose a novel first request first QoS (FRFQoS) entanglement routing strategy, which can optimize the application performance of each paired users in sequence and the overall efficiency of the network simultaneously.  

The channel allocation procedure with the FRFQoS strategy of the proposed DSER scheme can be detailed described with the following three steps:

\textbf{Step 1.} Calculate the matched channel vector $\rho_{ij}$ by the function $\rho_{ij}=Pair\_Ch(sp_i,sp_j)$. If $\rho_{ij}=\emptyset$, indicate that $(\nu_i,\nu_j)$ has no correlated wavelength channels, and then unlock the path if no active connections existed by the operation $V=Unlock(G,p_x,ch_x)$, where $x =i,j$. Meanwhile, put the request user pair $(\nu_i,\nu_j)$ into the queue $\textbf{Q}$ and set $lk_{ij}=0$. If $\rho_{ij}\neq \emptyset$, go to \textbf{Step 2}.

\begin{definition}
	$\rho_{ij}=Pair\_Ch(sp_i,sp_j)$. When at the same spectral distance from the central wavelength, energy conservation ensures that pairs of wavelengths are correlated\rm{~\cite{gayerTemperatureWavelengthDependent2008,joshiTrustedNodeFree2020, qi15userQuantumSecure2021}}. Given the available and occupied channel of user $\nu_i(\nu_j)$ on the path $p_i(p_j)$ is $c_i^a(c_j^a)$ and $c_i^o(c_j^o)$, match the entangled channel pairs using the distance criteria, which is equidistant from the pump frequency. 
\end{definition}

\begin{figure}
    \centering
    \includegraphics[width=0.6\linewidth]{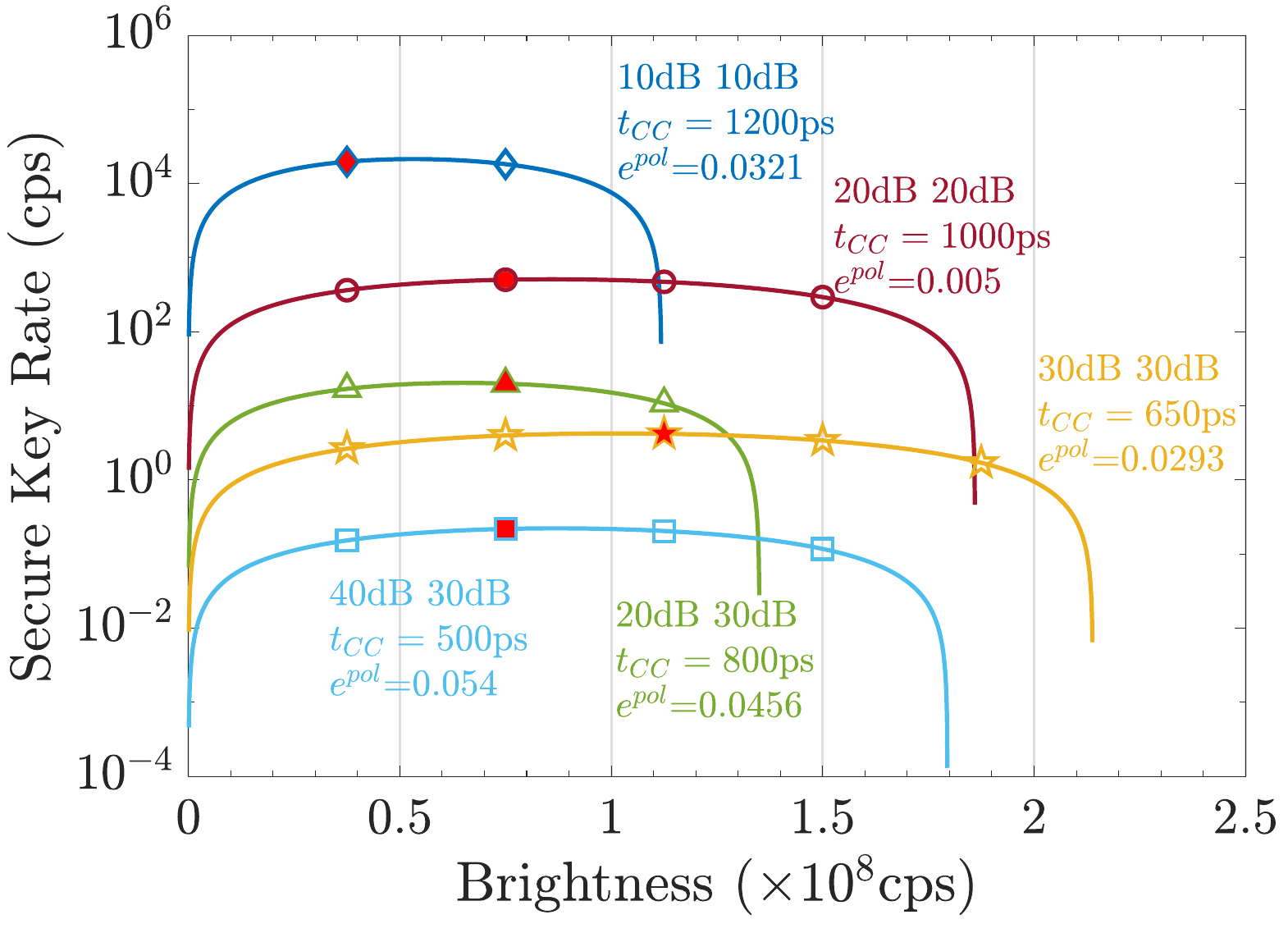}
    \caption{Secure key rate vs brightness $\mathrm{B}$ for different user loss $l_i,l_j$, coincidence window $t_{CC}$, and polarization measurement errors $e^{pol}$, dark count rate $\mathrm{DCR} = 300\mathrm{cps}$, and the brightness of each wavelength channel is $3.75\times 10^7\mathrm{cps}$. The red solid label indicates the optimal count of channels that achieve the highest secure key rate.}
    \label{pic:brightness_key}
\end{figure}

\textbf{Step 2.} Calculate the optimal count of channels for the requested user pair $(\nu_i, \nu_j)$ by the operation $\omega_{ij}=Optimal\_Ch(sp_i,sp_j)$, as the secure key rate are always firstly increased and then decreased versus the entangled photon source brightness, shown in Fig.~\ref{pic:brightness_key}. 

\begin{definition}
    \label{de:optimal}
	$\omega_{ij}=Optimal\_Ch(sp_i,sp_j)$. Calculate the optimal count of wavelength channel $\omega_{ij}$ using the QKD model given in the Appendix~\ref{chap: Analysis of the Secure Key Rate} \rm{~\cite{neumannModelOptimizingQuantum2021a}}.
\end{definition}

\begin{figure}[H]
    \centering
    \includegraphics[width=1\linewidth]{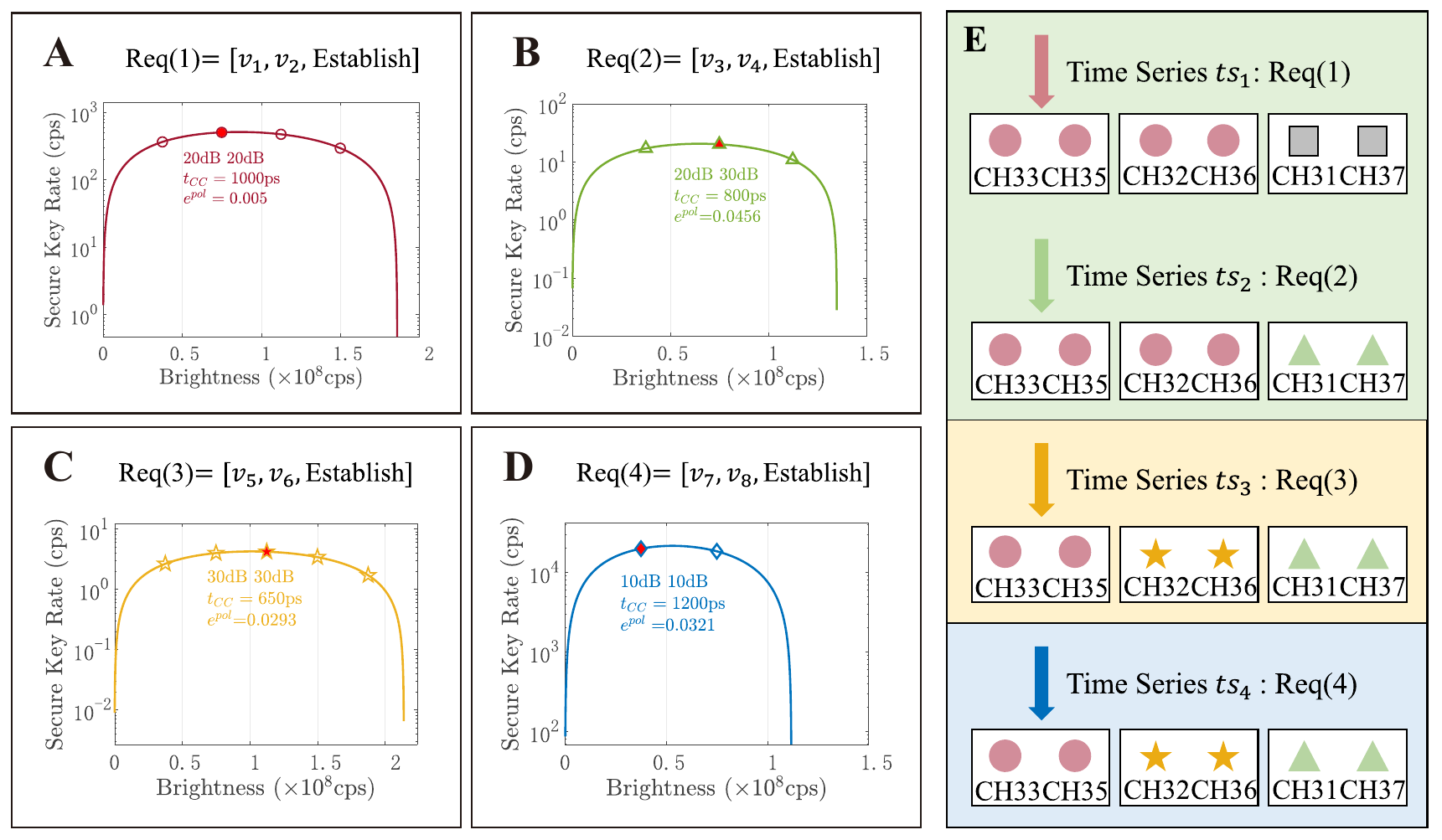}
    \caption{First request first QoS routing strategy. $ts_i$ means the request time series and $ts_{i+1}>ts_i$. $\mathrm{Req(n)}=[\nu_i,\nu_j,opt]$ means the user request, $n=1,2,3,4$. $t_{\mathrm{CC}}$ is the coincidence window, $e^{pol}$ is the individual polarization error probability and dark count rate $\mathrm{DCR}=300\mathrm{cps}$. The solid red label indicates the optimal channel count in (A)-(D). (E) indicates the channel allocated result.}
    \label{pic:fourcases}
\end{figure}

\textbf{Step 3.} Allocate the wavelength channels with FRFQoS strategy for requested user pair $(\nu_i, \nu_j)$. Here, the FRFQoS strategy can be divided into four different cases, according to $\rho=\rho_{ij} \cap (c_i^a+c_j^a)$, as shown in Fig.~\ref{pic:fourcases}. 

\begin{itemize}
    \item \textbf{Case 1: Satisfaction}.\\
    In this case, with the condition of $|\rho|\geq \omega_{ij}$, the DSER scheme can satisfy the connection construction request of the user pair $(\nu_i, \nu_j)$. Allocate the optimal channel count $\omega_{ij}$ to the user pair $(\nu_i, \nu_j)$, update the lowest-loss path set $\textbf{SP}$ and the channel allocation set $\textbf{SC}$, as shown in Fig.~\ref{pic:fourcases}(E) with time series $ts_1$. Then, lock the paired channels of the internal ports for the WDM forwarding devices on the path $p_i$ and $p_j$ by the operation $V=Lock(G,p_x,ch_x)$, where $x =i,j$.

    \item \textbf{Case 2: Partial}.\\
    In this case, with the condition of $0<|\rho| < \omega_{ij}$, the DSER scheme can only partially satisfy the connection construction request of the user pair $(\nu_i, \nu_j)$. Allocate the current available channel vector $\rho$ to user pair $(\nu_i, \nu_j)$ and add them into the QoS priority queue $\textbf{Q}$ with set $lk_{ij}=1$, as shown in Fig.~\ref{pic:fourcases}(E) with  time series $ts_2$. Then, lock the paired channels and update the set $\textbf{SP}$, $\textbf{SC}$ by the operation $V=Lock(G,p_x,ch_x)$, where $x =i,j$.
    
    \item \textbf{Case 3: Competition}.\\
    In this case, $|\rho| = 0$ and meanwhile with the condition of $\exists sc_x, sc_y\in $\rm{\textbf{SC}} where $(sc_x[2]=sc_y[2]~and~sc_x[3]=sc_y[3]) |(sc_x[1]\in\rho_{ij})$ or $(sc_x[2]=sc_y[3]~and~sc_x[3]=sc_y[2]) |(sc_x[1]\in\rho_{ij})$, find out the paired user nodes ($\nu_q$,$\nu_r$) with maximum request timetag value, which has surplus available channels. Then, the DSER scheme can allocate a channel pair to the requested pair ($\nu_i,\nu_j$) from the resources occupied by ($\nu_q$,$\nu_r$), as shown in Fig.~\ref{pic:fourcases}(E) with time series $ts_3$. Afterwards, both user pairs are added to the queue $\textbf{Q}$ and set $lk_{ij}=1, lk_{qr}=1$. Thirdly, unlock the competed channels by the operation $V=Unlock(G,p_x,ch_x)$, where $x =q,r$. Fourthly, lock the paired channel by the operation $V=Lock(G,p_x,ch_x)$, where $x =i,j$. Finally, update the set $\textbf{SP}$, $\textbf{SC}$.

    \item \textbf{Case 4: Infeasible}.\\
    In this case, $|\rho| = 0$ and meanwhile with the condition of $\forall sc_x, sc_y\in $\rm{\textbf{SC}} where $(sc_x[2]\neq sc_y[2]~and~sc_x[3]\neq sc_y[3])|(sc_x[1]\in\rho_{ij})$ or $(sc_x[2]\neq sc_y[3]~and~sc_x[3]\neq sc_y[2]) |(sc_x[1]\in\rho_{ij})$,
    the DSER scheme can not allocate any wavelength channels to the user pair $(\nu_i, \nu_j)$, as shown in Fig.~\ref{pic:fourcases}(E) with time series $ts_4$. Put $(\nu_i, \nu_j)$ into the QoS priority queue $\textbf{Q}$ and set $lk_{ij}=0$. Meanwhile, unlock the path $p_i$ by the operation $V=Unlock_{port}(G,p_i)$ and update the set $\textbf{SP}$.
\end{itemize}

Therefore, the proposed DSER scheme can be detailed described as Algorithm~\ref{al:DSER}, where the tensor-based lowest loss path searching procedure is defined as function $(G,\textbf{SP})=Path\_Search(G,s,\nu_x)$,
the channel allocation calculation with FRFQoS routing strategy is defined as function $(G,\textbf{SP},\textbf{SC},\textbf{Q})=FRFQoS(G,\textbf{SP},\textbf{SC},\textbf{Q})$ and the channel allocation reconfiguration procedure is defined as function $(G,\textbf{SP},\textbf{SC},\textbf{Q})=Del(G,\textbf{SP},\textbf{SC},\textbf{Q})$. Also, the calculation of the entanglement flow table $\textbf{D}=(d_1,d_2,\cdots,d_{n_4})$ is defined function $\textbf{D}=Calculate(\textbf{SP},\textbf{SC})$, where $n_4$ is the size of $\textbf{D}$ (see the Appendix~\ref{chap: Entanglement flow table}).

\renewcommand{\thealgorithm}{1}
\begin{algorithm}[H]
    \caption{the Differentiated service entanglement routing scheme} 
    \begin{algorithmic}[1] 
        \Require User request $\mathrm{Req}=\left [\nu_i,\nu_j,opt\right ]$ and network status database: $G, \textbf{SP}, \textbf{SC}, \textbf{Q}$.
        \Ensure Entanglement flow table $\textbf{D}$.
        \If{$opt=\mathrm{Establish}$}
            \State //Tensor-based Lowest Loss Path Searching Module
            \If{$\nu_i\notin \textbf{SP}$} $(G,\textbf{SP})=Path\_Search(G,s,\nu_i);$
            \EndIf
            \If{$\nu_j\notin \textbf{SP}$} $(G,\textbf{SP})=Path\_Search(G,s,\nu_j);$
            \EndIf
            \State //Channel Allocation Module with FRFQoS Routing Strategy
            \State Match $\rho_{ij}=Pair\_Ch(sp_i,sp_j);$
            \State Calculate $\omega_{ij} = Optimal\_Ch(sp_i,sp_j);$
            \State $(G,\textbf{SP},\textbf{SC},\textbf{Q})=FRFQoS(G,\textbf{SP},\textbf{SC},\textbf{Q});$
            \State \Return $\textbf{D}=Calculate(\textbf{SP},\textbf{SC})$.
        \ElsIf{$opt=\mathrm{Delete}$}
            \qquad \State //Channel Allocation Reconfiguration Module
            \qquad \For{$(i=1;i\leq n_2;i++)$}
                \qquad \If{$sc_i[2][3] = (\nu_i,\nu_j)$ or $sc_i[2][3] =(\nu_j,\nu_i)$}
                \qquad \State Delete $sc_i$ from $\textbf{SC}$;
                \qquad \EndIf
            \qquad \EndFor
            \qquad \If{$(\nu_i,\nu_j)\in \textbf{Q}$} Delete $(\nu_i,\nu_j)$ from $\textbf{Q}$;
            \qquad \EndIf
            \qquad \If{$\nu_i\notin \textbf{SC}[2]$} Update $\textbf{SP}$ and $G$;
            \qquad \EndIf
            \qquad \If{$\nu_j\notin \textbf{SC}[2]$} Update $\textbf{SP}$ and $G$;
            \qquad \EndIf
            \qquad \State $(G,\textbf{SP},\textbf{SC},\textbf{Q})=Del(G,\textbf{SP},\textbf{SC},\textbf{Q});$
            \qquad \State \Return $\textbf{D}=Calculate(\textbf{SP},\textbf{SC})$.
        \EndIf
    \end{algorithmic}
    \label{al:DSER}
\end{algorithm}

\section{Evaluation}
\label{chap:Evaluation}

In the evaluation, a continuous wave polarization-entangled photon source is used to generate the entangled photons, where a $775.06$ nm pump down-converts in a type-0 (MgO:PPLN) crystal to produce signal and idler photons. The central wavelength of the entangled photons is $1550.12$ nm, corresponding to the center of ITU channel $34$~\cite{tangDemonstration75Kmfiber2023}. 
Suppose that each pair of wavelength channels has been compensated for dispersion and has the same brightness in ITU Grid C-Band (100 GHz Spacing) standard. 
The detailed parameters performed in our evaluation are given in Table~\ref{parameter}.

\begin{table}[ht]
  \centering
  \caption{\label{parameter}The parameters used in the experimental evaluation. $\mathrm{B_{ch}}$ is the brightness of each wavelength channel pair. $t_{\mathrm{CC}}$ is the coincidence window and $\eta^{t_{\mathrm{CC}}}$ is the coincidence-window dependent detection efficiency. $e^{pol}$ is the individual polarization error probability. $\mathrm{DCR}$ is the dark count rate. $loss_\mathrm{user}$ is each user's internal optical device loss.}
  \setlength{\tabcolsep}{7mm}
  \renewcommand{\arraystretch}{1.5}
  \footnotesize
  \begin{tabular}{cccccc}
  \hline
  \boldmath $\mathrm{B_{ch}}$ & \boldmath $t_{CC}$ & \boldmath $\eta^{t_{CC}}$ & \boldmath $e^{pol}$ & \boldmath $\mathrm{DCR}$ & \boldmath $loss_{user}$\\
  \hline
  $3.75\times 10^7 \mathrm{cps}$ & $1000\mathrm{ps}$ & $0.761$ & $0.5\%$ & $300\mathrm{cps}$ & $2\mathrm{dB}$\\
  \hline
  \end{tabular}\\
\end{table}
\normalsize

\subsection{Evaluation on T/W-DM networks}
\label{chap:Evaluation on T/W-DM networks}

\begin{figure}
 \centering
 \includegraphics[width=1\linewidth]{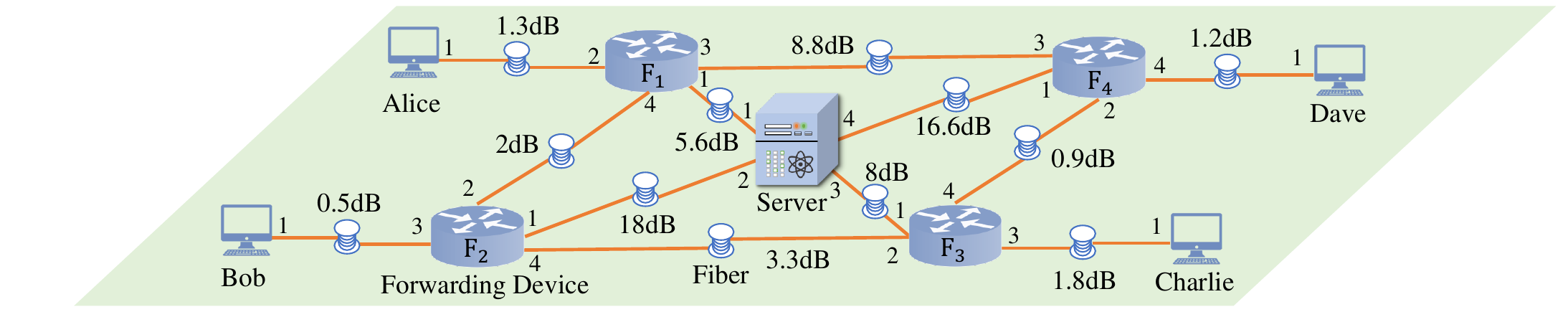}
 \caption{The experimental topology of TDM and WDM technology networks. The entangled photon source is connected to a WSS or a DWDM as a quantum server $S$, which allocates the wavelength channel to users. $\mathrm{F}_1, \mathrm{F}_2, \mathrm{F}_3, \mathrm{F}_4$ are four forwarding devices, and the number around the node represents the port of the device. $Alice, Bob, Charlie$ and $Dave$ are users.}
 \label{topology}
\end{figure}

The DSER scheme is evaluated on the topology in Fig.~\ref{topology}. The entangled photon source is connected to a wavelength selective switch (WSS) or a DWDM. $\mathrm{F}_1, \mathrm{F}_2, \mathrm{F}_3, \mathrm{F}_4$ are four routing forwarding devices, and the number around the node represent the port of the device. $Alice, Bob, Charlie$ and $Dave$ are users. 

\begin{table}[ht]
    \caption{\label{case}Different configurations of forwarding devices. $(1/2in,3/4out)$ indicates that ports $1$ and $2$ of the device are input ports, and ports $3$ and $4$ are output ports. The optical switches(OS), all-pass optical switches(AOS), DWDM, and WSS insertion losses are $0.5dB, 1dB, 1.5dB, 4dB$, respectively. The OS, AOS and WSS can support $\mathrm{CH}30, \mathrm{CH}31, \mathrm{CH}32, \mathrm{CH}33, \mathrm{CH}35, \mathrm{CH}36, \mathrm{CH}37$ and $\mathrm{CH}38$. The $1$-th port of the $1\times 3$ DWDM are the input ports, and the $2$-th to $4$-th port are output ports which support $\mathrm{CH}33, \mathrm{CH}35,$ and $\mathrm{CH}36$ respectively. The $1$-th port of the $1\times 4$ DWDM is the input ports, and the $2$-th to $5$-th port are output ports which support $\mathrm{CH}32, \mathrm{CH}33, \mathrm{CH}35,$ and $\mathrm{CH}36$ respectively.}
    \footnotesize
    \centering
    \setlength{\tabcolsep}{4mm}
    \renewcommand{\arraystretch}{1.85}
    \begin{tabular}{ccccccc}
    \hline
    \textbf{Case} & \textbf{\makecell{Network \\ Type}} & \boldmath $S$ & \boldmath $\mathrm{F}_1$ & \boldmath $\mathrm{F}_2$ & \boldmath $\mathrm{F}_3$ & \boldmath $\mathrm{F}_4$\\
    \hline
    $1$ & TDM & \makecell{Source + $5$ ports \\ WSS} & \makecell{$2\times 2$ \\ AOS} & \makecell{$2\times 2$ \\ AOS} & \makecell{$2\times 2$ \\ AOS} & \makecell{$2\times 2$ \\ AOS}\\
    $2$ & TDM & \makecell{Source + $5$ ports \\ WSS} & \makecell{OS($1/4$in, \\ $2/3$out)} & \makecell{OS($1/4$in, \\ $2/3$out)} & \makecell{OS($1/2$in, \\ $3/4$out)} & \makecell{OS($1/2$in, \\ $3/4$out)}\\
    $3$ & TDM+WDM & \makecell{Source + $5$ ports \\ WSS} & \makecell{$5$ ports \\ WSS} & \makecell{$5$ ports \\ WSS} & \makecell{$5$ ports \\ WSS} & \makecell{$5$ ports \\ WSS}\\
    $4$ & TDM+WDM & \makecell{Source + $5$ ports \\ WSS} & \makecell{$5$ ports \\ WSS} & \makecell{OS($1/4$in, \\ $2/3$out)} & \makecell{$2\times 2$ \\ AOS} & \makecell{$2\times 2$ \\ AOS}\\
    $5$ & TDM+WDM & \makecell{Source + $1\times 4$ \\ DWDM} & \makecell{$1\times 3$ \\ DWDM} & \makecell{$2\times 2$ \\ AOS} & \makecell{$2\times 2$ \\ AOS} & \makecell{$2\times 2$ \\ AOS}\\
    \hline
    \end{tabular}\\
\end{table}
\normalsize

Due to inherent bandwidth limitations, pure WDM quantum networks cannot support dynamic configuration. Therefore, the article focuses on TDM networks and the combination of TDM with WDM networks, with simulations their performance. The simulation involves the deployment of various forwarding devices on $S, \mathrm{F}_1, \mathrm{F}_2, \mathrm{F}_3, \mathrm{F}_4$ in Table~\ref{case}.

\begin{table}
  \caption{\label{result1} The evaluation results obtained by the proposed DSER. $l_A(l_B)$ indicates that the path loss between the server and the user $Alice(Bob)$. Channel$_{AB}$ indicates the entangled wavelength channel pair configured. $R_{AB}$ indicates the secure key rate generated by the user pair $(Alice,Bob)$. $A\mathrm{F}_2\mathrm{F}_1S\mathrm{F}_1\mathrm{F}_4\mathrm{F}_3B$ means the path $A\leftarrow \mathrm{F}_2\leftarrow \mathrm{F}_1\leftarrow S\rightarrow \mathrm{F}_1\rightarrow \mathrm{F}_4\rightarrow \mathrm{F}_3\rightarrow B$.}
  \footnotesize
  \setlength{\tabcolsep}{2mm}
  \renewcommand{\arraystretch}{1.5}
  \begin{tabular}{ccccccc}
    \hline
    \textbf{Case} & \textbf{Path} & \boldmath $l_A(\mathrm{dB})$ & \boldmath $l_B(\mathrm{dB})$ & \textbf{Channel}$_{AB}$ & \boldmath $R_{AB}(\mathrm{cps})$ & \textbf{Li's} \boldmath $R_{AB}(\mathrm{cps})$\\
    \hline
    $1$ & $A\mathrm{F}_1S\mathrm{F}_3\mathrm{F}_2B$ & $11.9$ & $17.8$ & $[\mathrm{CH}33, \mathrm{CH}35], [\mathrm{CH}32, \mathrm{CH}36]$ & $5398.8$ & $3897.2$\\
    $2$ & $A\mathrm{F}_1S\mathrm{F}_2B$ & $11.4$ & $23$ & $[\mathrm{CH}33, \mathrm{CH}35], [\mathrm{CH}32, \mathrm{CH}36]$ & $1828.1$ & $1319.8$\\
    $3$ & $A\mathrm{F}_1S\mathrm{F}_1\mathrm{F}_2B$ & $14.9$ & $20.1$ & $[\mathrm{CH}33, \mathrm{CH}35], [\mathrm{CH}32, \mathrm{CH}36]$ & $1592.8$ & $1149.9$\\
    $4$ & $A\mathrm{F}_1S\mathrm{F}_3\mathrm{F}_2B$ & $14.9$ & $17.3$ & $[\mathrm{CH}33, \mathrm{CH}35], [\mathrm{CH}32, \mathrm{CH}36]$ & $3035.8$ & $2191.5$\\
    $5$ & $A\mathrm{F}_1\mathrm{F}_2\mathrm{F}_3S\mathrm{F}_2B$ & $19.6$ & $21$ & $[\mathrm{CH}33, \mathrm{CH}35]$ & $316.581$ & $316.581$\\
    \hline
  \end{tabular}\\
\end{table}
\normalsize

The proposed DSER scheme search the lowest loss path and optimal channel vector between the user pair $(Alice,Bob)$ under different network configurations and the results are presented in Table~\ref{result1}. Compared with Li \textit{et al.} first request first server (FRFS) scheme~\cite{liFirstRequestFirst2022}, the generated key rate is significantly improved in the Case $1-4$. In the Case $5$, the network has only a channel pair to establish entanglement. The evaluation results show that our DSER scheme can support quantum networks with multiple infrastructures.

\subsection{Evaluation with Dynamic Request Sequences of Users}
\label{chap:Evaluation with Dynamic Request Sequences of Users}
\subsubsection{The Lowest Loss Paths}
Our proposed DSER scheme searches for the lowest loss path in the current network status when the user dynamically accesses, but the network has limited path resources, and the user who first requests can lock the path node port.

\begin{table}[H]
    \caption{\label{path_sequence}Two different request sequences. $ts_i$ means the request time series and $ts_i>ts_{i-1}$. The username represents the path with the lowest search loss for the current user.}
    \footnotesize
    \setlength{\tabcolsep}{9mm}
    \renewcommand{\arraystretch}{1.2}
    \begin{tabular}{ccccc}
        \hline
        \textbf{Request Sequence} & \boldmath $ts_1$ & \boldmath $ts_2$ & \boldmath $ts_3$ & \boldmath $ts_4$ \\
        \hline
        $1$ & $Alice$ & $Bob$ & $Charlie$ & $Dave$ \\
        $2$ & $Dave$ & $Charlie$ & $Bob$ & $Alice$ \\
        \hline
    \end{tabular}\\
\end{table}
\normalsize

\begin{figure}[ht]
    \centering
    \includegraphics[width=1\linewidth]{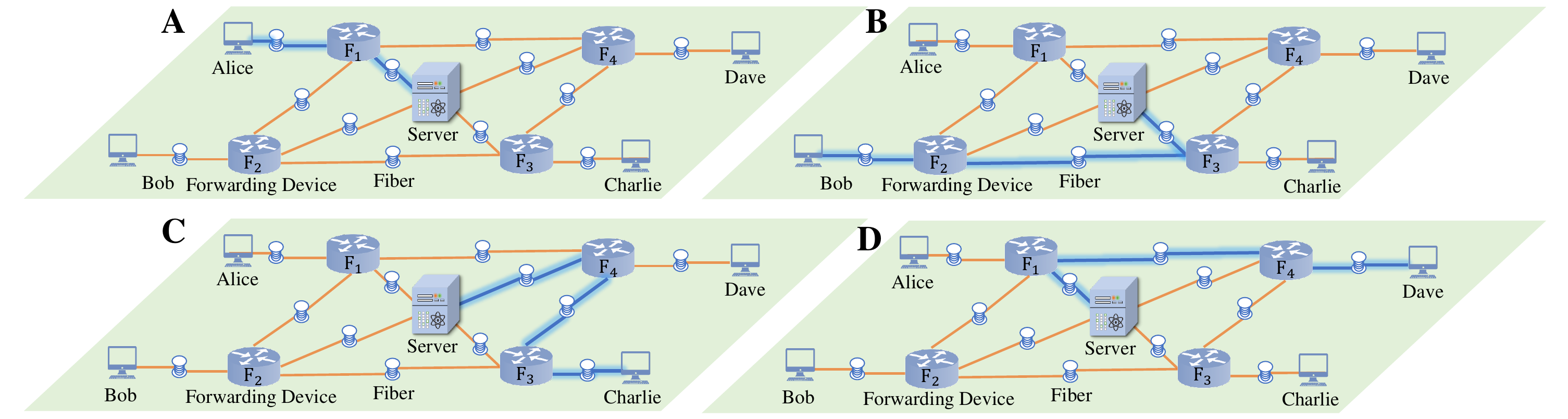}
    \caption{The user request sequence: $Alice \rightarrow Bob \rightarrow Charlie \rightarrow Dave$. The network configuration is Case $4$ in Table~\ref{case}, the server $S$ is the entangled photon source + $5$ ports WSS, $\mathrm{F}_1,\mathrm{F}_2,\mathrm{F}_3,\mathrm{F}_4$ represent the $5$ ports WSS, OS$(1/4in,2/3out)$, $2\times 2$ AOS, $2\times 2$ AOS and $Alice,Bob,Charlie,Dave$ represent the users. The blue line indicates the lowest loss path from the server $S$ to the user.}
    \label{path1}
\end{figure}

\begin{figure}[ht]
    \centering
    \includegraphics[width=1\linewidth]{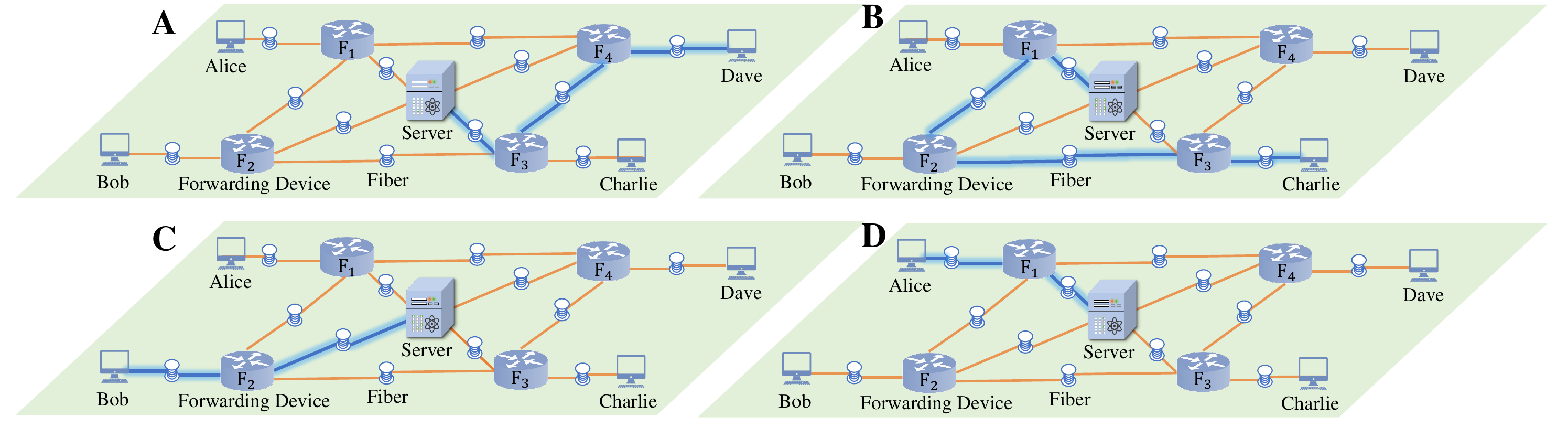}
    \caption{The user request sequence: $Dave \rightarrow Charlie \rightarrow Bob \rightarrow Alice$. The network configuration is Case $4$ in Table~\ref{case}, the server $S$ is the entangled photon source + $5$ ports WSS, $\mathrm{F}_1,\mathrm{F}_2,\mathrm{F}_3,\mathrm{F}_4$ represent the $5$ ports WSS, OS$(1/4in,2/3out)$, $2\times 2$ AOS, $2\times 2$ AOS and $Alice,Bob,Charlie,Dave$ represent the users. The blue line indicates the lowest loss path from the server $S$ to the user.}
    \label{path2}
\end{figure}

As can be seen from Fig.~\ref{path1} and Fig.~\ref{path2}, based on the network configuration of Case $4$ in Table.~\ref{case}, the lowest loss path between the quantum server and the user varies with the request sequence in Table.~\ref{path}.
When multiple users request a path, the one who makes the request first can reserve the specific node port on the lowest loss path. This means that other users will need to search for a different path that is available on the remaining resources. This DSER scheme helps to efficiently allocate network resources and prevent conflicts among multiple users. It also ensures that the first user requesting the path has priority.

\begin{table}[H]
    \caption{\label{path}The lowest loss path and loss value under different dynamic request sequences of users. Sequence is shown in Table.~\ref{path_sequence}. $S\mathrm{F}_1\mathrm{F}_2A$ means the path $S\rightarrow \mathrm{F}_1\rightarrow \mathrm{F}_2\rightarrow A$.}
    \footnotesize
    \setlength{\tabcolsep}{6mm}
    \renewcommand{\arraystretch}{1.2}
    \begin{tabular}{|c|c|cccc|}
        \hline
        \textbf{Sequence} & \textbf{Result} & \textbf{Alice} & \textbf{Bob} & \textbf{Charlie} & \textbf{Dave}\\
      \hline
      \multirow{2}*{1} & \textbf{Path} & $S\mathrm{F}_1A$ & $S\mathrm{F}_3\mathrm{F}_2B$ &$S\mathrm{F}_4\mathrm{F}_3C$ & $S\mathrm{F}_1\mathrm{F}_4D$ \\
      \cline{2-6}
		~ &\textbf{Loss(dB)} & $14.9$ & $17.3$ & $25.3$ & $24.6$\\
        \hline
      \multirow{2}*{2} & \textbf{Path} & $S\mathrm{F}_1A$ & $S\mathrm{F}_2B$ & $S\mathrm{F}_1\mathrm{F}_2\mathrm{F}_3C$ & $S\mathrm{F}_3\mathrm{F}_4D$ \\
      \cline{2-6}
		~ &\textbf{Loss(dB)} & $14.9$ & $23$ & $22.2$& $16.1$\\
      \hline
    \end{tabular}\\
\end{table}
\normalsize

\subsubsection{The Allocation of Wavelength Channels}
The DSER scheme can dynamically allocate the optimal wavelength channel pair under different dynamic request sequences of Users. Evaluate the DSER scheme based on the network configuration of Case $4$ in Table.~\ref{case} with two different request sequences in Table~\ref{sequence}.

\begin{table}[H]
  \caption{\label{sequence}Two different request sequences. $ts_i$ means the request time series and $ts_i>ts_{i-1}$. $+(A,B)$ means the user pair establish a connection, $-(A,B)$ means the user pair delete a connection.}
  \footnotesize
  \setlength{\tabcolsep}{4.2mm}
  \renewcommand{\arraystretch}{1.2}
  \begin{tabular}{ccccccc}
    \hline
    \textbf{Request Sequence} & \boldmath $ts_1$ & \boldmath $ts_2$ & \boldmath $ts_3$ & \boldmath $ts_4$ & \boldmath $ts_5$ & \boldmath $ts_6$\\
    \hline
    $1$ & $+(A,B)$ & $+(A,C)$ & $+(A,D)$ & $+(B,C)$ & $-(A,B)$ & $+(C,D)$\\
    $2$ & $+(A,D)$ & $+(A,C)$ & $+(A,B)$ & $-(A,D)$ & $+(B,C)$ & $+(B,D)$\\
    \hline
  \end{tabular}\\
\end{table}
\normalsize

\begin{figure}[H]
    \centering
    \includegraphics[width=1\linewidth]{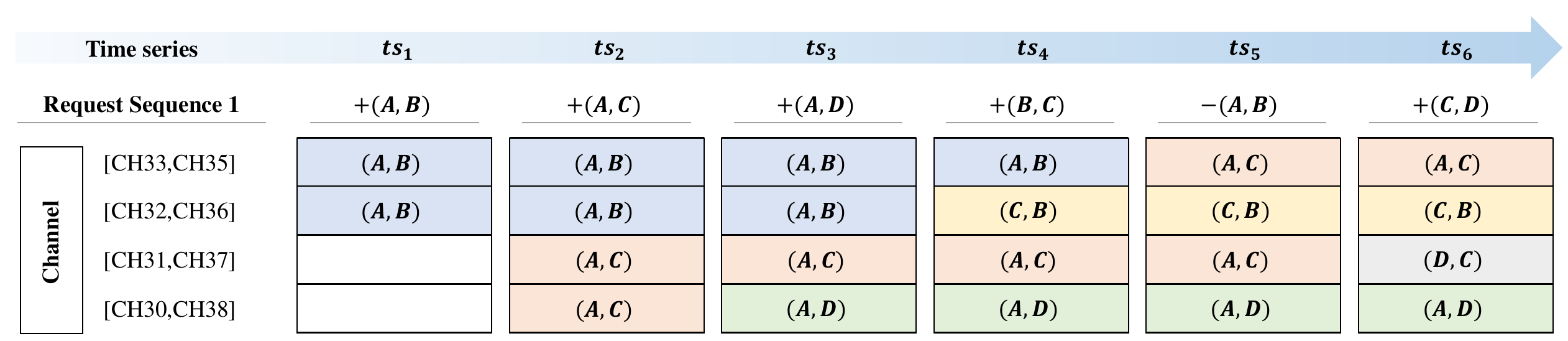}
    \caption{The channel allocation under the request sequence $1$ in Table~\ref{sequence}, $ts_i$ means the request time series and $ts_i>ts_{i-1}$. The sequence of user pairs corresponds to channel resources and $[ CH33, CH35] \rightarrow (A,B)$ indicates $CH33\rightarrow A, CH35\rightarrow B$.}
    \label{sequence1}
   \end{figure}

\begin{figure}[H]
    \centering
    \includegraphics[width=1\linewidth]{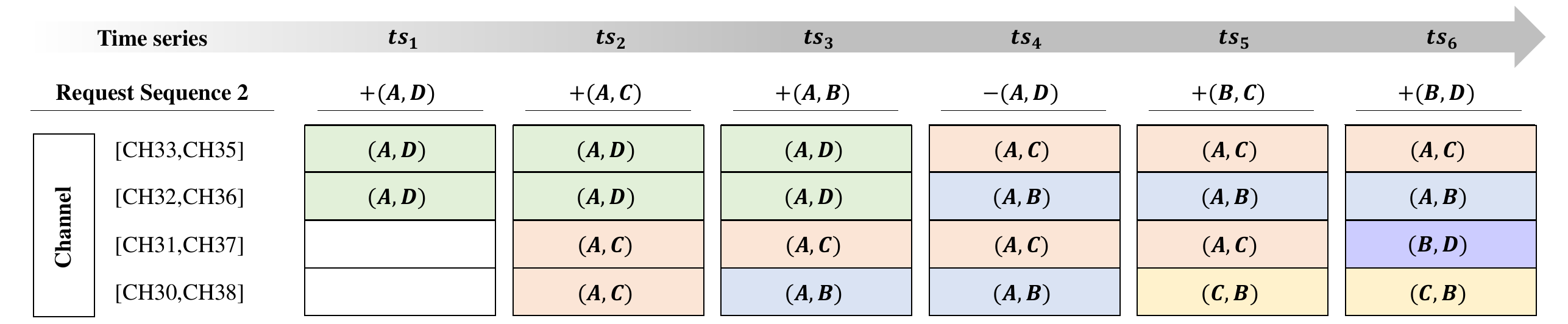}
    \caption{The channel allocation under the request sequence $2$ in Table~\ref{sequence}, $ts_i$ the request time series and $ts_i>ts_{i-1}$. The sequence of user pairs corresponds to channel resources and $[ CH33, CH35] \rightarrow (A,B)$ indicates $CH33\rightarrow A, CH35\rightarrow B$.}
    \label{sequence2}
\end{figure}

The allocation of wavelength channels is depicted in Fig.~\ref{sequence1} for the first request sequence and in Fig.~\ref{sequence2} for the second request sequence. When a user pair establishes a connection in different request sequences, it is possible for the wavelength channel pairs allocated to the same user pair to vary. The DSER scheme searches the lowest loss path in the current network status and allocates the wavelength channel pairs with the FRFQoS routing strategy, providing more entanglement channel resources for users who request earlier. 

\begin{figure}
    \centering
    \includegraphics[width=1\linewidth]{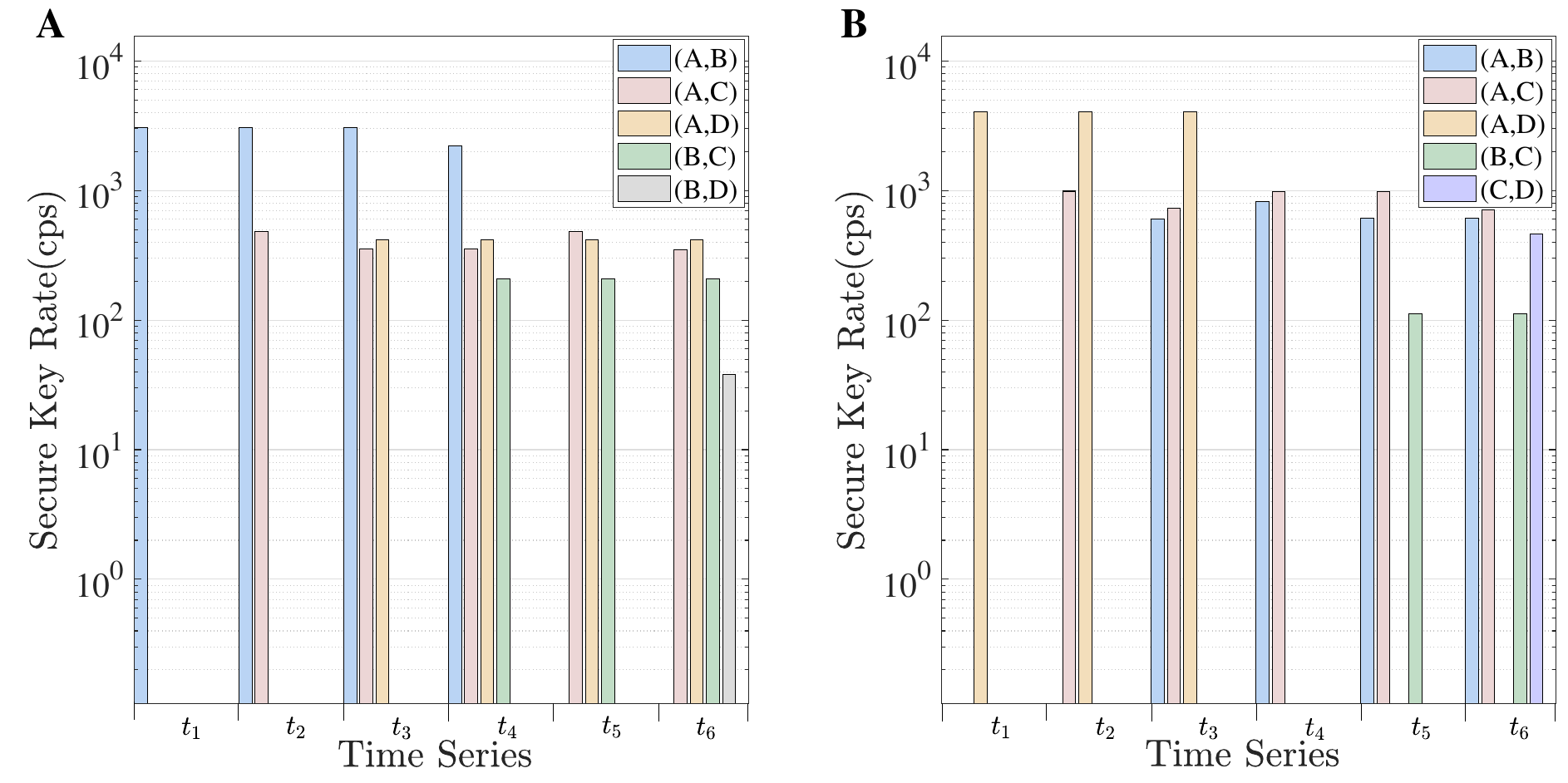}
    \caption{The secure key rate of the user pair under different dynamic request sequences, (A) the first request sequence scenario, (B) the second request sequence scenario. $A,B,C,D$ represent the users $Alice, Bob, Charlie, Dave$ and the $ts_i,i=1,2,\cdots,6$ represent the time series and $ts_{i+1}>ts_i$.}
    \label{sequence_rate}
\end{figure}

Fig.~\ref{sequence_rate} shows that the secure key rates with wavelength channel allocation under different dynamic request sequences. Fig.~\ref{sequence_rate}(A) shows the secure key rates of each time series for the first request sequence scenario. The first user pair $(A,B)$ has two channel pairs in Fig.~\ref{sequence_rate}(A) with time series $ts_1$, and the secure key rate is $\mathrm{3025.8cps}$. Add the second user pair $(A,C)$ with the secure key rate of $\mathrm{480.4473cps}$, the channel resources are fully utilized. As the third user pair $(A,D)$ joins, the channel resources for the user pair $(A,C)$ reduce to a channel pair, while the third user pair $(A,D)$ obtains a channel pair with the secure key rate of $\mathrm{407.6546cps}$. During this process, the secure key rate for the user pair $(A,C)$ exhibits a downward trend, dropping to $\mathrm{346.8943cps}$. Moreover, add the fourth user pair $(B,C)$ with the secure key rate of $\mathrm{199.5970cps}$, the secure key rate for the user pair $(A,B)$ decreases to $\mathrm{2191.5cps}$. At the time series $ts_5$, the user pair $(A,B)$ delete the entanglement connection, and their channel resources are allocated to QoS priority queued users. This reallocation leads to a partial increase in the secure key rate for the user pair $(A,C)$ to $\mathrm{480.4473cps}$. Finally, the surplus channel resources for the user pair $(A,C)$ was completed to the new request user pair $(C,D)$ with the time series $ts_6$, the secure key rate for the user pair $(C,D)$ is $\mathrm{37.1287cps}$.

Fig.~\ref{sequence_rate}(B) shows the secure key rates of each time series for the second request sequence scenario. The first user pair $(A,D)$ and second user pair $(A,C)$ has two channel pairs with time series $ts_1$, $ts_2$, and their secure key rate are $\mathrm{4002.3cps}$ and $\mathrm{981.8165cps}$, respectively. This indicates the full utilization of channel resources within the network. The secure key rate for the user pair $(A,C)$ decrease to $\mathrm{708.7990cps}$ with the addition of the third user pair $(A,B)$, and the third user pair $(A,B)$ obtains a channel pair with the secure key rate of $\mathrm{589.4675cps}$.
With the user pair $(A,D)$ deletes the connection in the time series $ts_4$, the secure key rate for the user pair $(A,B)$ and $(A,C)$ increased to $\mathrm{816.4990cps}$ and $\mathrm{981.8165cps}$ respectively. 
At the time series $ts_5$ and $ts_6$, the new request user pair $(B,C)$ and $(B,D)$ complete the surplus channel resources for the previous users, the secure key rate for the user pair $(A,B)$ and $(A,C)$ increased to $\mathrm{589.4675cps}$ and $\mathrm{708.7990cps}$ respectively. Simultaneously, the secure key rate for the new request user pair $(B,C)$ and $(B,D)$ are $\mathrm{109.6995cps}$ and $\mathrm{447.1377cps}$ respectively.

\section{Conclusion}

In this article, we propose a differentiated service entanglement routing scheme for configurable quantum entanglement distribution networks. The proposed DSER scheme firstly finds out the lowest loss paths and supported wavelength channels between the entangled photon source with the user with the tensor-based lowest loss path searching algorithm, and then allocates the optimal paired channels with the differentiated routing strategies.
The evaluation results show that the proposed DSER scheme can achieve the maximization of the number of entanglement connections and the overall network efficiency under different network configurations and user dynamic request sequences.
The proposed DSER scheme can be applied to large-scale, high-topology complex quantum entanglement distribution networks, and the next step can be to study quantum memory and quantum repeater to provide users with more diverse quality of service guarantees.











\printbibliography

\appendix
\section*{Appendix}
\renewcommand\thefigure{A\arabic{figure}}
\renewcommand\thedefinition{A\arabic{definition}}
\renewcommand\thetable{A\arabic{table}}
\setcounter{figure}{0}
\setcounter{table}{0}

\section{Modeling of the network}
\label{chap: Modeling of the network}

\subsection{Tensor}
\label{chap: Tensor}

\begin{definition}
    Tensor. A tensor is a multidimensional array. In general, tensors of order 1 are called vectors, tensors of order 2 are called matrices, and tensors of order 3 and above are called tensors. For a tensor $\mathcal{X}\in \mathbb{R}^{I_1\times I_2\times \cdots\times I_N}$, $N$ is the order of the tensor, also known as an array or mode, where $I_i$ is called the dimension of the $i$ order. The $(i_1, i_2, \cdots, i_n)$ element of $\mathcal{X}$ is represented as $\mathcal{X}_{i_1, i_2, \cdots, i_n}, i_n\in \{ 1, 2, \cdots, I_n \}, 1\le n\le N$.
\end{definition}

\begin{figure}[H]
 \centering
 \includegraphics[width=0.7\linewidth]{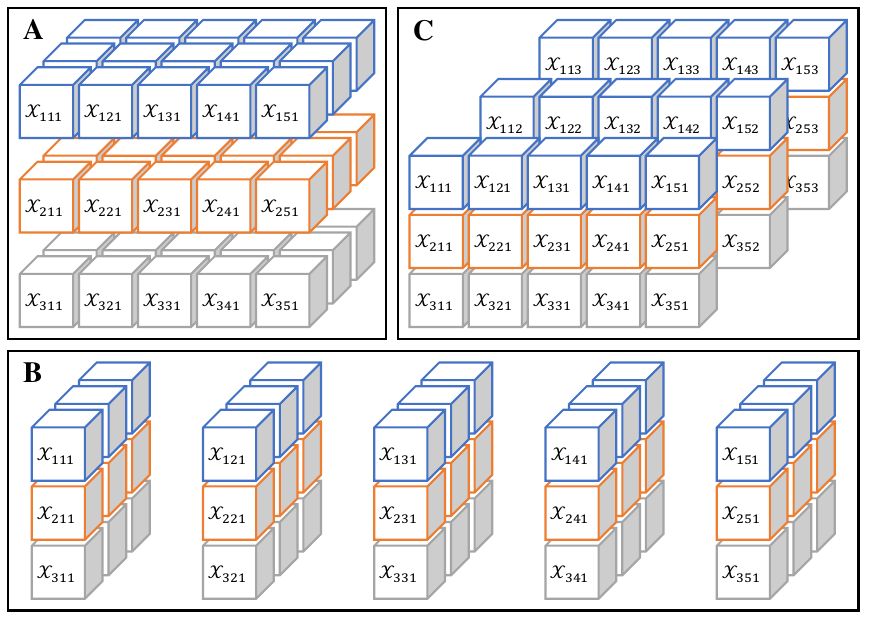}
 \caption{(A) Horizontal slices $X_{i::}$. (B) Lateral slices $X_{:j:}$. (C) Frontal slices $X_{::k}$}.
 \label{pic: slice}
\end{figure}

\begin{definition}
    Tensor slicing. A slice is a two-dimensional portion of a tensor defined by fixing all but two subscripts. As shown in \rm{Fig.~\ref{pic: slice}}, using the three order tensor $\mathcal{X}\in \mathbb{R}^{I\times J\times K}$, there are three slicing methods, namely, horizontal slices are represented by symbol $X_{i::}$, lateral slices are represented by symbol $X_{:j:}$, and frontal slices are represented by symbol $X_{::k}$.
\end{definition}

\subsection{Modeling of the network edges}
\label{chap: Modeling of the network edges}

The edge $e_{\alpha\beta}^{ij}$ represents a quantum link that sends quantum information from $\alpha$-th port of the node $\nu_i$ to $\beta$-th port of the node $\nu_j$. 

\begin{equation}
 e_{\alpha\beta}^{ij}=\left(i,\alpha,j,\beta,c_{\alpha\beta}^{ij}\right),
 \label{e} 
\end{equation}

Where $e_{\alpha\beta}^{ij}[\gamma]$ represents the $\gamma$-th element, $\gamma\in\{1,2,3,4,5\}$. The $e_{\alpha\beta}^{ij}[1], e_{\alpha\beta}^{ij}[3]$ is the node index, $1\leq i, j\leq k$. The $e_{\alpha\beta}^{ij}[2], e_{\alpha\beta}^{ij}[4]$ is the node port index, $1\leq \alpha \leq m_i$ and $1\leq \beta \leq m_j$, $m_i$($m_j$) are the total counts of ports for node $\nu_i(\nu_j)$. And $e_{\alpha\beta}^{ij}[5]=c_{\alpha\beta}^{ij}$ is the path loss value.
If there is no path between the $\alpha$-th port of node $\nu_i$ and the $\beta$-th port of node $\nu_j$, $c_{\alpha\beta}^{ij}=\infty$.
Thus, the path loss matrix from node $\nu_i$ to node $\nu_j$ can be defined as
\begin{equation}
	C_{ij} = \begin{bmatrix}
		c^{ij}_{11} & \cdots & c^{ij}_{1m_j}\\ 
		\vdots & \ddots & \vdots\\ 
		c^{ij}_{m_i1} & \cdots & c^{ij}_{m_im_j}
	\end{bmatrix}.
 \label{Cij} 
\end{equation}

\begin{figure}[H]
 \centering
 \includegraphics[width=1\linewidth]{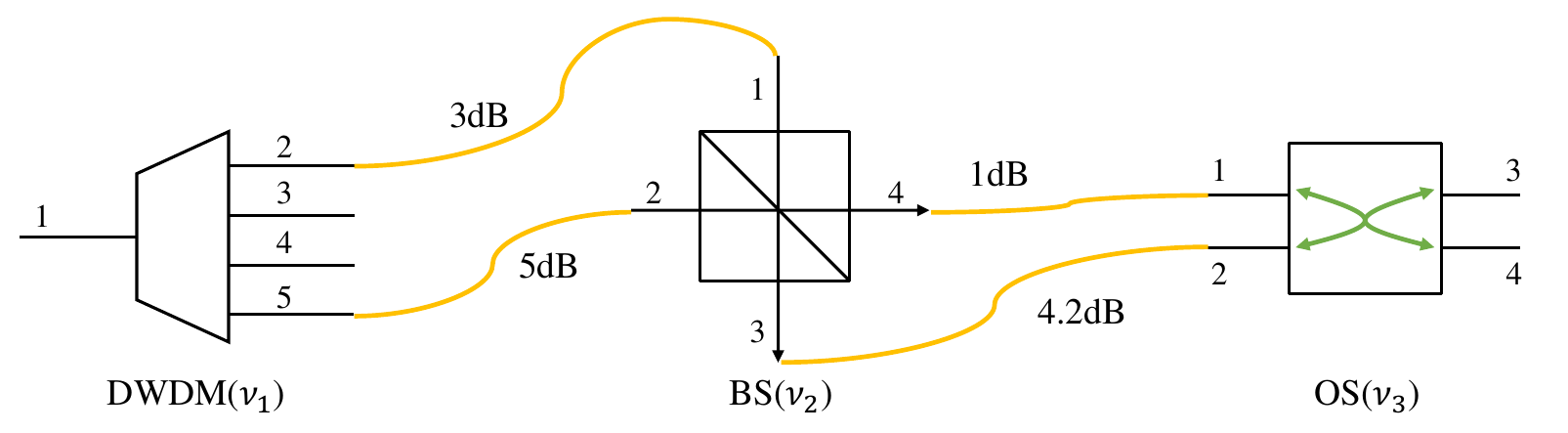}
 \caption{Example of edges between DWDM, BS, and OS}
 \label{pic:edges}
\end{figure}

Fig.~\ref{pic:edges} shows the edges between DWDM($\nu_1$), beam splitters(BS)($\nu_2$) and OS($\nu_3$), where port $2$ of DWDM is connected to port $1$ of BS with a path loss of $3dB$: $e_{21}^{12}=\left(1,2,2,1,3\right)$. Similarly, port $5$ of DWDM is connected to port $2$ of BS with a path loss of $5dB$: $e_{52}^{12}=\left(1,5,2,2,5\right)$. And port $4$ of BS is connected to port $1$ of OS with a path loss of $1dB$: $e_{41}^{23}=\left(2,4,3,1,1\right)$. Similarly, port $3$ of BS is connected to port $2$ of OS with a path loss of $4.2dB$: $e_{32}^{23}=\left(2,3,3,2,4.2\right)$. 

\subsection{Modeling of the network nodes}
\label{chap: Modeling of the network nodes}

Each node $\nu_i$ contains a pass-through loss tensor and a support channel vector, defined as a bi-variate vector $\nu_i\triangleq(\mathcal{W}_i,T_i), 1\leq i\leq k$. $\mathcal{W}_i\in \mathbb{R}^{m_i\times m_i\times |T_i|}$, $m_i$ is the total counts of ports for node $\nu_i$ and $T_i$ is the supported wavelength channel vector. Considering the different forwarding devices in a quantum network, the $T_i$ for node $\nu_i$ is defined as 

\begin{equation}
    T_i=\begin{cases}[-3], & \mathrm{Entangled ~photon ~source}\\ [-2], & \mathrm{User}\\ [-1], & \mathrm{BS}\\ [0], & \mathrm{OS~or~AOS} \\ \left[\mathrm{ch}_1,\mathrm{ch}_2,\dots,\mathrm{ch}_{|T_i|}\right], & \mathrm{DWDM~or~WSS}\end{cases}.
\end{equation}

The entangled photon source, user and TDW devices, set $|T_i|=1$, which means these nodes can support all the wavelengths. The WDM devices set $T_i=\left[\mathrm{ch}_1,\mathrm{ch}_2,\dots,\mathrm{ch}_{|T_i|}\right]$, which means the allowed wavelength channel vector.

\begin{figure}[ht]
    \centering
    \includegraphics[width=0.8\linewidth]{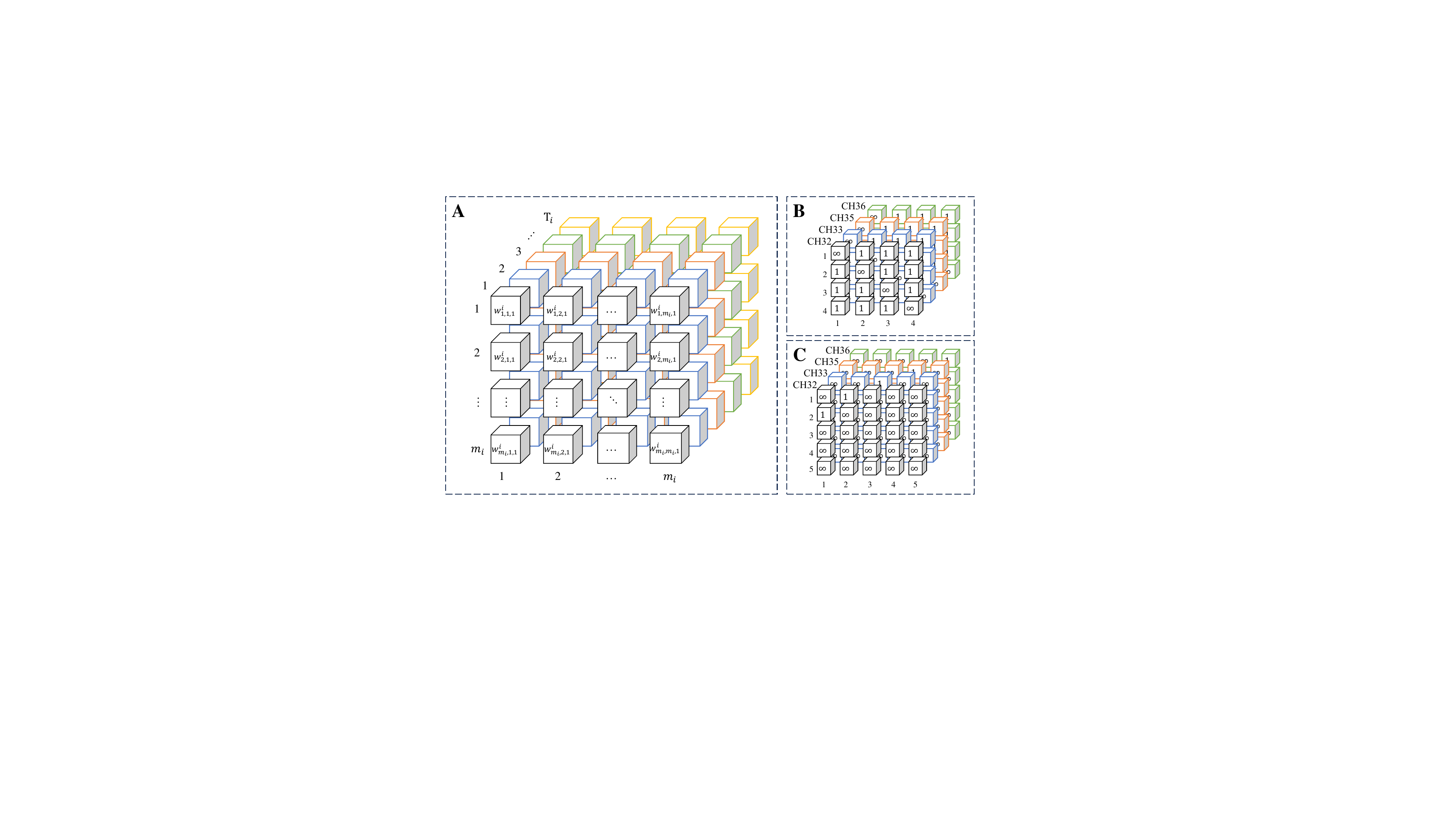}
    \caption{Modeling of the network nodes. (A) Node pass-through loss tensor $\mathcal{W}_i\in \mathbb{R}^{m_i\times m_i\times |T_i|}$. $m_i$ indicates the count of ports for the node $\nu_i$, and $|T_i|$ indicates the count of supported channels for the node $\nu_i$. Each element $w^{i}_{j,k,l}$ is the pass-through loss from $j$-th port to $k$-th port of the $l$-th wavelength channel for the node $\nu_i$, $1\leq j, k \leq m_i$ and $1\leq l\leq |T_i|$. (B) shows the pass-through loss tensor of 4-port WSS, where $m_{WSS}=4$, $T_{WSS}=[\mathrm{CH}32,\mathrm{CH}33,\mathrm{CH}35,\mathrm{CH}36]$ and $|T_i|=4$. (C) shows the pass-through loss tensor of 5-port DWDM, where $m_{DWDM}=5$, $T_{DWDM}=[\mathrm{CH}32,\mathrm{CH}33,\mathrm{CH}35,\mathrm{CH}36]$ and $|T_i|=5$.}
    \label{pic: nodes}
\end{figure}

As shown in Fig.~\ref{pic: nodes}(A), the pass-through loss tensor of the node $\nu_i$ is a three-order tensor $\mathcal{W}_i\in\mathbb{R}^{m_i\times m_i\times |T_i|}$, where $m_i$ represents the count of ports for the node $\nu_i$ and $T_i$ represents the supported wavelength channel vector for node $\nu_i$. Each element $w^i_{j,k,l}$ indicates the pass-through loss from $j$-th port to $k$-th port in $l$-th wavelength channel for the node $\nu_i$, $1\leq j,k\leq m_i, 1\leq l\leq |T_i|$. If $j$-th port cannot be reached from $k$-th port, $W_{j,k,:} = \infty$. Fig.~\ref{pic: nodes}(B)-(C) shows $1\times 4$ WSS and $1\times 5$ DWDM pass-through loss tensor.

\subsection{Lock and Unlock Operations}
\label{chap:lock and unlock}

In order to describe the DSER scheme clearly, the definitions of lock and unlock the node port and channel are given as follows:

\begin{definition}
	$V=Lock_{port}(G,p_x)$. For the arbitrary adjacent ports $a$ and $b$ for TDM devices ($|T_x|=1$) in the path $p_x$($a\leftrightarrow b$), get the pass-through loss element $\mathrm{w}^x_{z,z}=(y_1,y_2)$, where $z=\{a,b\}$. $y_1$ is the locked port number($\infty$ means no lock), $y_2$ is the count of users who are also using this locked path. Update the loss tensor by setting $\mathrm{w}^x_{a,a} = (b,y_2+1)$ and $\mathrm{w}^x_{b,b} = (a,y_2+1)$.
    \label{def:lockport}
\end{definition}

\begin{definition}
	$V=Unlock_{port}(G,p_x)$. For the arbitrary adjacent ports $a$ and $b$ for TDM devices ($T_x=0$) in the path $p_x$($a\leftrightarrow b$), get the pass-through loss element $\mathrm{w}^x_{z,z,1}=(y_1,y_2)$, where $z=\{a,b\}$. $y_1$ is the locked port number($\infty$ means no lock), $y_2$ is the count of users who are also using this locked path. If $y_2-1=0$, update the loss tensor by setting $\mathrm{w}^x_{a,a,1} = (\infty,0)$ and $\mathrm{w}^x_{b,b,1} = (\infty,0)$. Else, update the loss tensor by setting $\mathrm{w}^x_{a,a,1} = (b,y_2-1)$ and $\mathrm{w}^x_{b,b,1} = (a,y_2-1)$.
    \label{def:unlockport}
\end{definition}

\begin{definition}
	$V=Lock(G,p_x,ch_x)$. For the arbitrary adjacent ports $a$ and $b$ for WDM devices ($T_x>0$) in the path $p_x$($a\leftrightarrow b$), get the pass-through loss vector $\mathrm{W}^x_{z,z,ch_x}$, where $z=\{a,b\}$. Update the loss vector by setting $\mathrm{w}^q_{a,a,ch_x} = b$ and $\mathrm{w}^q_{b,b,ch_x} = a$.
    \label{def:lock}
\end{definition}

\begin{definition}
	$V=Unlock(G,p_x,ch_x)$. For the arbitrary adjacent ports $a$ and $b$ for WDM devices ($T_x>0$) in the path $p_x$($a\leftrightarrow b$), get the pass-through loss vector $\mathrm{W}^x_{z,z,ch_x}$, where $z=\{a,b\}$. Update the loss vector by setting $\mathrm{w}^q_{a,a,ch_x} = \infty$ and $\mathrm{w}^q_{b,b,ch_x} = \infty$.
    \label{def:unlock}
\end{definition}

\section{Analysis of the Secure Key Rate}
\label{chap: Analysis of the Secure Key Rate}

This section presents an overview of bi-photon entanglement distribution for evaluating the performance between distant communication users (node $\nu_i$ and $\nu_j$).
Assume the polarized continuous-wave pumped entangled photon source is used for node $\nu_i$ and $\nu_j$. The brightness of the photon source is $B$, and the overall efficiency between the source and node $\nu_i$ ($\nu_j$) is $\eta_i$ ($\eta_j$). Thus, the measured single photon counts for node $\nu_i$ and $\nu_j$ can be calculated by 
\begin{equation}
 S_{\mathrm{i}}^{\mathrm{m}}=B \eta_{\mathrm{A}}+\mathrm{DC}_{\mathrm{i}} \text { and } S_{\mathrm{j}}^{\mathrm{m}}=B \eta_{\mathrm{B}}+\mathrm{DC}_{\mathrm{j}},
\label{eq2}
\end{equation}
where $\mathrm{DC}_{\mathrm{i}}$ and $\mathrm{DC}_{\mathrm{j}}$ are the dark counts.

The total measured coincidences count between node $\nu_i$ and $\nu_j$ $\mathrm{CC}^{\mathrm{m}}$ is defined as
\begin{equation}
 \mathrm{CC}^{\mathrm{m}}=\eta^{t_{\mathrm{CC}}} \mathrm{CC}^{\mathrm{t}}+\mathrm{CC}^{\mathrm{acc}}
\label{cc_m},
\end{equation}
where $t_{\mathrm{CC}}$ is the coincidence window, $\eta^{t_{\mathrm{CC}}}$ is the coincidence-window dependent detection efficiency. $\mathrm{CC}^{\mathrm{t}}$ is the true coincident counts, which represents the rate of the events that two photons of a pair must be detected to observe their polarization correlation. $\mathrm{CC}^{\mathrm{acc}}$ is the accidental coincidence count. 

The true coincidence count rate is calculated as 
\begin{equation}
 \mathrm{CC}^{\mathrm{t}}=B \eta_{\mathrm{i}} \eta_{\mathrm{j}}
\label{CC_t}.
\end{equation}

The count rate of accidental coincidences in coincidence window $t_{\mathrm{cc}}$ is given by
\begin{equation} 
 \mathrm{CC}^{\mathrm{acc}}=S_{\mathrm{i}}^{\mathrm{m}} S_{\mathrm{j}}^{\mathrm{m}}t_{\mathrm{CC}} 
 \label{CC_acc}.
\end{equation}

The count rate of the erroneous coincidences $\mathrm{CC}^{\mathrm{err}}$ is calculated as
\begin{equation}
 \mathrm{CC}^{\mathrm{err}}=\eta^{t_{\mathrm{CC}}} \mathrm{CC}^{\mathrm{t}} e^{\mathrm{pol}}+\frac{1}{2} \mathrm{CC}^{\mathrm{acc}}
\label{CC_err},
\end{equation}
where $e^{\mathrm{pol}}$ is the system polarization error probability.

Therefore, the quantum bit error rate (QBER $ER$) can be expressed by
\begin{equation}
ER=\frac{\mathrm{CC}^{\mathrm{err}}}{\mathrm{CC}^{\mathrm{m}}}
\label{E}=\frac{\eta^{t_{\mathrm{CC}}} \mathrm{CC}^{\mathrm{t}} e^{\mathrm{pol}}+\frac{1}{2} \mathrm{CC}^{\mathrm{acc}}}{\eta^{t_{\mathrm{CC}}} \mathrm{CC}^{\mathrm{t}}+\mathrm{CC}^{\mathrm{acc}}}.
\end{equation}

This article assumes that node $\nu_i$ and node $\nu_j$ perform the BBM92 protocol to generate the information-theoretical-secure keys.
Since the noise parameters are independent of measurement settings, assume the phase error equals bit error $ER$. Thus, the secure key rate ($R^{\mathrm{s}}$) can be expressed as
\begin{equation}
 R^{\mathrm{s}}=\frac{1}{2} \mathrm{CC}^{\mathrm{m}}\left[1-f\left(ER\right) \mathrm{H}_{2}(ER) - \mathrm{H}_{2}(ER)\right]
\label{eq4},
\end{equation}
where $f(ER)$ is the information reconciliation efficiency and $ \mathrm{H}_2(x)$ is the binary Shannon entropy, which is defined as:
\begin{equation}
    \mathrm{H}_{2}(x)=-x \log _{2}(x)-(1-x) \log _{2}(1-x)
\label{H2}.
\end{equation}

\section{Entanglement flow table}
\label{chap: Entanglement flow table}
The network status database will update the network graph $G$, the lowest loss path set $\textbf{SP}$, the channel allocation set $\textbf{SC}$ and the QoS priority queue $\textbf{Q}$ after the module. 
Combined the case 4 in Table~\ref{case} and the request sequence 1 in Table~\ref{sequence}, get the each user's path $p_x, x=\{ Alice, Bob, Charlie, Dave \}$ in the lowest loss path set $\textbf{SP}$ and used channel vector in the channel allocation set $\textbf{SC}$.

\begin{table}[H]
    \caption{\label{path_sequence1_flow}Path $p_x, x=\{ Alice, Bob, Charlie, Dave \}$ in the lowest loss path set $\textbf{SP}$ and occupied channel vector in the channel allocation set $\textbf{SC}$.}
    \footnotesize
    \setlength{\tabcolsep}{13mm}
    \renewcommand{\arraystretch}{1.5}
    \begin{tabular}{c|c|c}
        \hline
        \textbf{User} & \textbf{Path} & \textbf{Channel}\\
      \hline
      $Alice$ & $(0,1,1,1,1,2,5,1)$ & $\mathrm{CH}35,\mathrm{CH}38$\\
      \hline
      $Bob$ & $(0,3,3,1,3,2,2,4,2,3,6,1)$ & $\mathrm{CH}36$\\
      \hline
      $Charlie$ & $(0,4,4,1,4,2,3,4,3,3,7,1)$ & $\mathrm{CH}32,\mathrm{CH}33,\mathrm{CH}37$\\
      \hline
      $Dave$ & $(0,1,1,1,1,3,4,3,4,4,8,1)$ & $\mathrm{CH}30,\mathrm{CH}31$\\
      \hline
    \end{tabular}\\
\end{table}
\normalsize

The odd digit value of path is the node index, and the even digit value is the port index.
$Path=(0,1,1,1,1,2,5,1)$ represents the $1$-th port of server $0$ to $1$-th port on node $1$ and pass through the node $1$ to port $2$, finally reaches the $1$-th port of node $5$. The end node of $p_x$ is the user node $\nu_x$. We can match the used channel for the user $\nu_x$ in the set $\textbf{SC}$. If the match is successful, the network controller will deploy entanglement flow table for each forwarding device. Give the example flow table of forwarding devices in Table~\ref{flowtable_S}, Table~\ref{flowtable_1} and Table~\ref{flowtable_3}, where the server has no input port mode and OS has no channel selection.

However, the set $\textbf{SC}$ is not updated in the fourth case of FRFQoS routing strategy, the network status database will store the user pair in $\textbf{Q}$, and deploy the flow table after the $\textbf{SP}$ and $\textbf{SC}$ are successfully matched.

\begin{table}[H]
    \caption{\label{flowtable_S}Example of entanglement flow table $d_0$ for server $S$.}
    \footnotesize
    \setlength{\tabcolsep}{15.5mm}
    \renewcommand{\arraystretch}{1.5}
    \begin{tabular}{ccc}
        \hline
        \textbf{Rule} & \textbf{Output port} & \textbf{Channel} \\
      \hline
      $\mathrm{R}1$ & $1$ & $\mathrm{CH}30,\mathrm{CH}31,\mathrm{CH}35,\mathrm{CH}38$\\
      \hline
      $\mathrm{R}2$ & $3$ & $\mathrm{CH}36$\\
      \hline
      $\mathrm{R}3$ & $4$ & $\mathrm{CH}32,\mathrm{CH}33,\mathrm{CH}37$\\
      \hline
    \end{tabular}\\
\end{table}
\normalsize

\begin{table}[H]
    \caption{\label{flowtable_1}Example of entanglement flow table $d_1$ for forwarding device $F_1$ (WSS).}
    \footnotesize
    \setlength{\tabcolsep}{12mm}
    \renewcommand{\arraystretch}{1.5}
    \begin{tabular}{cccc}
        \hline
        \textbf{Rule} & \textbf{Input port} & \textbf{Output port} & \textbf{Channel}\\
      \hline
      $\mathrm{R}1$ & $1$ & $2$ & $\mathrm{CH}35,\mathrm{CH}38$\\
      \hline
      $\mathrm{R}2$ & $1$ & $3$ & $\mathrm{CH}30,\mathrm{CH}31$\\
      \hline
    \end{tabular}\\
\end{table}
\normalsize

\begin{table}[H]
    \caption{\label{flowtable_3}Example of entanglement flow table $d_3$ for forwarding device $F_3$ (OS).}
    \footnotesize
    \setlength{\tabcolsep}{18.5mm}
    \renewcommand{\arraystretch}{1.5}
    \begin{tabular}{ccc}
        \hline
        \textbf{Rule} & \textbf{Input port} & \textbf{Output port}\\
      \hline
      $\mathrm{R}1$ & $1$ & $2$\\
      \hline
    \end{tabular}\\
\end{table}
\normalsize
\end{document}